\documentclass[prd,nofootinbib,aps,superscriptaddress,tightenlines,preprintnumbers]{revtex4}

\usepackage{latexsym,amsmath,amssymb,graphicx,hyperref}

\def\a{\alpha}		\def\b{\beta}		
\def\d{\delta}		\def\e{\epsilon}		
\def\g{\gamma}		\def\h{\eta}		
					\def\l{\lambda}
\def\m{\mu}		\def\n{\nu}			\def\o{\omega}
					\def\r{\rho}
\def\s{\sigma}

\def\bnew{\beta_{*}}
\def\aether{{\ae}ther}
\def\lag{{\mathcal{L}}}
\def\ham{{\cal{H}}}

\def\kv{\vec{k}}
\def\pd{\partial}
\def\bA{\bar{A}}
\def\ba{\bar{a}}
\def\be{\begin{equation}}
\def\ee{\end{equation}}
\begin{document}
\preprint{CALT-68-2712}
\title{Instabilities in the {\AE}ther}
\author{Sean M. Carroll, Timothy R. Dulaney, Moira I. Gresham, and Heywood Tam} 
\email[]{seancarroll@gmail.com}
\email[]{dulaney@theory.caltech.edu}
\email[]{moira@theory.caltech.edu}
\email[]{tam@theory.caltech.edu}
\affiliation{California Institute of Technology,~Pasadena, CA 91125, USA}
\date{\today}

\begin{abstract}
We investigate the stability of theories in which Lorentz invariance is spontaneously
broken by fixed-norm vector ``\aether'' fields.  
Models with generic kinetic terms are plagued either by 
ghosts or by tachyons, and are therefore physically unacceptable.  There are precisely three
kinetic terms that are not manifestly unstable:  a sigma model $(\partial_\mu A_\nu)^2$, 
the Maxwell Lagrangian $F_{\mu\nu}F^{\mu\nu}$,  and a scalar Lagrangian
$(\partial_\mu A^\mu)^2$.   The timelike sigma-model case
is well-defined and stable when the vector norm is fixed by a constraint; however, when
it is determined by minimizing a potential there is necessarily a tachyonic ghost, and therefore
an instability.  In the Maxwell and scalar cases, the Hamiltonian is unbounded below,
but at the level of
perturbation theory there are fewer degrees of freedom and the models are stable.  However, in these two theories 
there are obstacles to smooth evolution for certain choices of initial data.

\end{abstract}
\preprint{}
\maketitle

\section{Introduction}

The idea of spontaneous violation of Lorentz invariance through tensor fields
with non-vanishing expectation values has garnered substantial attention in 
recent years \cite{Will:1972zz, Gasperini:1987nq, Kostelecky:1989jw, Colladay:1998fq, Jacobson:2000xp, Eling:2003rd, Carroll:2004ai, Jacobson:2004ts,Lim:2004js, Eling:2004dk,Dulaney:2008ph, Jimenez:2008sq}.
Hypothetical interactions between Standard Model fields and Lorentz-violating (LV)
tensor fields are tightly constrained by a wide variety of experimental probes,
in some cases leading to limits at or above the Planck scale \cite{Colladay:1998fq, Kostelecky:2000mm, Carroll:2004ai, Elliott:2005va, Mattingly:2005re, Will:2005va, Jacobson:2008aj}. 

If these constraints are to be taken seriously, it is necessary to have a sensible theory
of the dynamics of the LV tensor fields themselves, at least at the level of 
low-energy effective field theory.  The most straightforward way to construct such a theory
is to follow the successful paradigm of scalar field theories with spontaneous
symmetry breaking, by introducing a tensor potential that is minimized at some
non-zero expectation value, in addition to a kinetic term for the fields.  (Alternatively, 
it can be a derivative of the field that obtains an expectation value, as in 
ghost condensation models \cite{ArkaniHamed:2003uy, ArkaniHamed:2005gu, Cheng:2006us}.)  As an additional simplification, we may 
consider models in which the nonzero expectation value is enforced by a 
Lagrange multiplier constraint, rather than by dynamically minimizing a potential;
this removes the ``longitudinal'' mode of the tensor from consideration, and may
be thought of as a limit of the potential as the mass near the minimum is 
taken to infinity.  In that case, there will be a vacuum manifold of zero-energy
tensor configurations, specified by the constraint.

All such models must confront the tricky question of stability.  
Ultimately, stability problems stem from the basic fact that the metric has an
indefinite signature in a Lorentzian spacetime.  Unlike in the case of scalar fields,
for tensors it is necessary to use the spacetime metric to define both the kinetic
and potential terms for the fields.  A generic choice of potential would
have field directions in which the energy is unbounded from below, leading to
tachyons, while a generic choice of kinetic term would have modes with negative
kinetic energies, leading to ghosts.   Both phenomena represent instabilities;
if the theory has tachyons, small perturbations grow exponentially in time at
the linearized level, while if the theory has ghosts, nonlinear interactions 
create an unlimited number of positive- and negative-energy excitations
\cite{Carroll:2003st}.  There is no simple argument that these
unwanted features are necessarily present in any model of LV tensor fields,
but the question clearly warrants careful study.

In this paper we revisit the question of the stability of theories of dynamical Lorentz
violation, and argue that most such theories are unstable.  In particular, we 
examine in detail the case of a vector field $A_\mu$ with a nonvanishing expectation
value, known as the ``{\ae}ther'' model or a ``bumblebee'' model.  For generic choices of
kinetic term, it is straightforward to show that the Hamiltonian of such a model is
unbounded from below, and there exist solutions with bounded initial data that grow exponentially in time.  

There are three specific
choices of kinetic term for which the analysis is more subtle.  These are the 
sigma-model kinetic term,
\be
  {\cal{L}}_K = -\frac{1}{2} \partial_\mu A_\nu \partial^\mu A^\nu\,,
  \label{sigmamodel}
\ee
which amounts to a set of four scalar fields defined on a target space with a Minkowski metric;
the Maxwell kinetic term,
\be
  {\cal L}_K = -\frac{1}{4}F_{\mu\nu}F^{\mu\nu}\,,
  \label{maxwell}
\ee
where $F_{\mu\nu} = \partial_\mu A_\nu - \partial_\nu A_\mu$ is familiar from 
electromagnetism; and what we call the ``scalar'' kinetic term,
\be
  {\cal L}_K = \frac{1}{2} (\partial_\mu A^\mu)^2\,,
  \label{scalar}
\ee
featuring a single scalar degree of freedom.
Our findings may be summarized as follows:
\begin{itemize}
\item  The sigma-model Lagrangian with the vector field constrained by a Lagrange
multiplier to take on a timelike expectation value is the only \aether\ theory for which the
Hamiltonian is bounded from below in every frame, ensuring stability.     In a companion paper,
we examine the cosmological behavior and observational constraints on this
model \cite{Carroll:2009en}.  If the vector field is
spacelike, the Hamiltonian is unbounded and the model is unstable.
However, if the constraint in the sigma-model theory
is replaced by a smooth potential, allowing the length-changing mode to
become a propagating degree of freedom, that mode is necessarily ghostlike (negative
kinetic energy) and tachyonic (correct sign mass term), and the Hamiltonian is unbounded 
below, even in the timelike case.  
It is therefore unclear whether models of this form can arise in any full theory. 
\item In the Maxwell case, the Hamiltonian is unbounded below; however, a perturbative
analysis does not reveal any explicit instabilities in the form of tachyons or ghosts.
The timelike mode of the vector acts as a Lagrange multiplier, and there are fewer
propagating degrees of freedom at the linear level
(a ``spin-1'' mode propagates, but not a ``spin-0'' mode).
Nevertheless, singularities can arise in evolution from generic initial data: for a spacelike vector,
for example, the field evolves to a configuration in which the
fixed-norm constraint cannot be satisfied (or perhaps just to a point where the effective field theory breaks down).  In the timelike case, a certain subset of initial data
is well-behaved, but, provided the vector field couples only to conserved currents, the theory reduces precisely to conventional electromagnetism, with no
observable violations of Lorentz invariance.  It is unclear whether there exists a subset
of initial data that leads to observable violations of Lorentz invariance while 
avoiding problems in smooth time evolution.
\item The scalar case is superficially similar to the Maxwell case, in that the Hamiltonian
is unbounded below, but a perturbative analysis does not reveal any instabilities.
Again, there are fewer degrees of freedom at the linear 
level; in this case, the spin-1 mode does not propagate.
There is a scalar degree of freedom, but it does not correspond to a propagating
mode at the level of perturbation theory (the dispersion relation is conventional, but the
energy vanishes to quadratic order in the perturbations).
For the timelike \aether\ field, obstacles arise in the time evolution that are similar to
those of a spacelike vector in the Maxwell case; for a spacelike \aether\ field with
a scalar action, the behavior is less clear.
\item For any other choice of kinetic term, \aether\ theories are always unstable.
\end{itemize}
Interestingly, these three choices of \aether\ dynamics are precisely those for which
there is a unique propagation speed for all dynamical modes; this is the same
condition required to ensure that the Generalized Second Law is respected by a
Lorentz-violating theory \cite{Dubovsky:2006vk,Eling:2007qd}.

One reason why our findings concerning stability seem more restrictive than those
of some previous analyses is that we insist on perturbative stability in all Lorentz frames,
which is necessary in theories where the form of the Hamiltonian is frame-dependent.
In a Lorentz-invariant field theory, it suffices to pick a Lorentz frame and examine the
behavior of small fluctuations; if they grow exponentially, the model is unstable, while
if they oscillate, the model is stable.  In Lorentz-violating theories, in contrast, such
an analysis might miss an instability in one frame that is manifest at the linear level in some other 
frame \cite{Kostelecky:2001xz, Mattingly:2005re,Adams:2006sv}.  
This can be traced to the fact that a perturbation that is ``small'' in one
frame (the value of the perturbation is bounded everywhere along some initial
spacelike slice), but grows exponentially with time as measured in that frame,
will appear ``large'' (unbounded on every spacelike slice) in some other frame.

As an explicit example, consider a model of a timelike vector with a background
configuration $\bar{A}_\mu = (m, 0, 0, 0)$, and perturbations 
$\delta a^\mu = \epsilon^\mu e^{-i\omega t} e^{i\vec k \cdot \vec x}$, where $\epsilon^\mu$
is some constant polarization vector.  In this frame, we will see that
the dispersion relation takes the form
\be
  \omega^2 = v^2 \vec k^2\,.
\ee 
Clearly, the frequency $\omega$ will be real for every real wave vector $\vec k$, 
and such modes simply oscillate rather than growing in time.  It is tempting to conclude
that models of this form are perturbatively stable for any value of $v$.  
However, we will see below that when $v > 1$, there exist other frames (boosted with
respect to the original) in which $\vec k$ can be real but $\omega$ is necessarily
complex, indicating an instability.  
These correspond to wave vectors for which, evaluated in the original
frame, both $\omega$ and $\vec k$ are complex.  Modes with complex
spatial wave vectors are not considered to be ``perturbations,'' since
the fields blow up at spatial infinity.  However, in the presence of
Lorentz violation, a complex spatial wave vector in one frame may
correspond to a real spatial wave vector in a boosted frame.  We will
show that instabilities can arise from initial data defined on a
constant-time hypersurface (in a boosted frame) constructed solely
from modes with real spatial wave vectors.  Such modes are bounded at
spatial infinity (in that frame), and could be superimposed to form
wave packets with compact support.  Since the notion of stability is
not frame dependent, the existence of at least one such frame
indicates that the theory is unstable, even if there is no linear
instability in the \aether\ rest frame.

Several prior investigations have considered the question of stability in theories with
LV vector fields.  Lim \cite{Lim:2004js} calculated the Hamiltonian for small perturbations
around a constant timelike vector field in the rest frame, and derived restrictions on
the coefficients of the kinetic terms.  Bluhm et al. \cite{Bluhm:2008yt} also examined the 
timelike case with a Lagrange multiplier constraint, and showed that the Maxwell kinetic 
term led to stable dynamics on a certain branch of the solution space if the
vector was coupled to a conserved current.  It was also found, in \cite{Bluhm:2008yt}, that 
most LV vector field theories have Hamiltonians that are unbounded below.  Boundedness of the Hamiltonian was also considered in \cite{Chkareuli:2006yf}. In the context of effective field theory, Gripaios \cite{Gripaios:2004ms} analyzed small fluctuations of LV vector fields about a flat background. Dulaney, 
Gresham and Wise \cite{Dulaney:2008ph}  showed that only the Maxwell choice was stable to small perturbations in the spacelike case assuming the energy of the linearized modes was non-zero.\footnote{This effectively eliminates the scalar case.}  Elliot, Moore, and Stoica \cite{Elliott:2005va}  showed that the sigma-model kinetic term is stable in the presence of a constraint, but not with a potential.

In the next section, we define notation and fully specify the models we are considering.
We then turn to an analysis of the Hamiltonians for such models, and show that they are
always unbounded below unless the kinetic term takes on the sigma-model form and the vector field is timelike.
This result does not by itself indicate an instability, as there may not be any dynamical
degree of freedom that actually evolves along the unstable direction.  Therefore, in the following
section we look carefully at linear stability around constant configurations, and isolate
modes that grow exponentially with time.  In the section after that we show that the models
that are not already unstable at the linear level 
end up having ghosts, with the exception of the Maxwell and scalar cases. 
We then examine some features of those two theories in particular.

\section{Models}

We will consider a dynamical vector field $A_\mu$
propagating in Minkowski spacetime with signature $(-+++)$.  The action takes the form
\be
  S_A = \int d^4x \left({\cal L}_K + {\cal L}_V\right)\,,
\ee
where ${\cal L}_K$ is the kinetic Lagrange density and ${\cal L}_V$ is (minus) the potential.
A general kinetic term that is quadratic in derivatives of the field can be written\footnote{In terms of the coefficients, $c_i$, defined in \cite{Jacobson:2004ts} and used in many other publications on {\ae}ther theories, 
\be
\b_i = {c_i \over 16 \pi G m^2}   
\ee
where $G$ is the gravitational constant.}
\begin{equation}
\label{aLag}
  {\cal{L}}_K = -\b_1(\partial_\m A_\n)(\partial^\m A^\n) - \beta_2 (\partial_\m A^\m)^2 
  - \b_3 (\partial_\m A_\n)(\partial^\n A^\m) 
  -{\beta_4} {A^\m A^\n  \over m^2} (\partial_\m A_\rho)(\partial_\n A^\rho)\,.
\end{equation}
In flat spacetime, setting the fields to constant values at infinity, we can integrate by parts
to write an equivalent Lagrange density as 
\begin{equation}
\label{aLag2}
{\cal{L}}_K = -{1\over 2}\b_1  F_{\m\n}F^{\m\n} 
-\bnew(\partial_\m A^\m)^2 - {\beta_4 } {A^\m A^\n \over m^2} (\partial_\m A_\rho)(\partial_\n A^\rho)\,,
\end{equation}
where $F_{\m\n}  = \partial_\m A_\n - \partial_\n A_\m$ and we have defined
\be
  \bnew = \beta_1 + \beta_2 +\beta_3\,.
\ee
In terms of these variables, the models specified above with no linear instabilities or
negative-energy ghosts are:
\begin{itemize}
\item Sigma model:  $\b_1 = \bnew$,
\item Maxwell: $\bnew = 0$, and
\item Scalar: $\b_1 = 0$,
\end{itemize}
in all cases with $\b_4 = 0$.

The vector field will obtain a nonvanishing vacuum expectation value from the 
potential.  For most of the paper we will take the potential to be a Lagrange
multipler constraint that strictly fixes the norm of the vector:
\be
{\cal L}_V =  \lambda(A^{\mu} A_{\mu} \pm m^2)\,,
\label{constraintl}
\ee
where $\l$ is a Lagrange multiplier whose variation enforces the constraint 
\be
A^{\mu} A_{\mu} = \mp m^2\,.
\label{constraint}
\ee
If the upper sign is chosen, the vector will be timelike, and it will be spacelike for the
lower sign.  Later we will examine how things change when the constraint is replaced
by a smooth potential of the form ${\cal L}_V = - V(A_\mu) \propto \xi(A_\mu A^\mu \pm m^2)^2$.
It will turn out that the theory defined with a smooth potential is only stable in the limit as $\xi \rightarrow \infty$. In any case, unless we specify otherwise, we assume that the norm of the vector is determined by the constraint (\ref{constraint}).

We are left with an action
\be
\label{vf action}
  S_A = \int d^4x \left[-{1\over 2}\b_1  F_{\m\n}F^{\m\n} 
  - \bnew(\partial_\m  A^\m)^2 
  -{\beta_4} {A^\m A^\n  \over m^2} (\partial_\m A_\rho)(\partial_\n A^\rho)
  + \lambda(A^{\mu} A_{\mu} \pm m^2)\right]\,.
\ee
The Euler-Lagrange equation obtained by varying with respect to $A_\mu$ is
\be
  \label{ele1}
  \b_1\partial_\m F^{\m \n} + \bnew \partial^\n \partial_\m A^\m 
  + \beta_4 G^\n = -\lambda A^\n\,,
\ee
where we have defined
\be
\label{Gnu}
  G^\n = \frac{1}{m^2}\left[A^\lambda (\partial_\lambda A^\sigma)F_\sigma{}^\n
  +A^\sigma(\partial_\lambda A^\lambda \partial_\sigma A^\nu 
                      +A^\lambda \partial_\lambda \partial_\sigma A^\nu)\right]\,.
\ee
Since the fixed-norm condition (\ref{constraint}) is a constraint, we can consistently
plug it back into the equations of motion.  Multiplying (\ref{ele1}) by $A_\nu$ and
using the constraint, we can solve for the Lagrange multiplier,
\be
  \lambda = \pm\frac{1}{m^2}\left(\b_1 \partial_\m F^{\m \n} 
  + \bnew \partial^\n \partial_\m A^\m + \b_4 G^\n\right)A_\n \,.
\ee
Inserting this back into (\ref{ele1}), we can write the equation of motion as a system of
three independent equations:
\begin{align}
Q_\r \equiv \left(\eta_{\r \n} \pm {A_\r A_\n \over m^2}\right) 
  \left(\b_1\partial_\m F^{\m \n} + \bnew \partial^\n \partial_\m A^\m 
  + \b_4 G^\n\right) = 0.
  \label{eoms}
\end{align} 
The tensor $\eta_{\rho\nu} \pm m^{-2}A_\rho A_\nu$ acts to take what would be the equation
of motion in the absence of the constraint, and project it into the hyperplane orthogonal
to $A_\mu$.
There are only three independent equations because $A^\r Q_\r$ vanishes identically, given the fixed norm constraint.

\subsection{Validity of effective field theory}\label{Validity of effective field theory}

As in this paper we will restrict our attention to classical field theory, it is important to 
check that any purported instabilities are found in a regime where a low-energy
effective field theory should be valid.  The low-energy degrees of freedom in our
models are Goldstone bosons resulting from the breaking of Lorentz invariance.
The effective Lagrangian will consist of an infinite series of terms of progressively
higher order in derivatives of the fields, suppressed by appropriate powers of some
ultraviolet mass scale $M$.  If we were dealing with the theory of a scalar field
$\Phi$, the low-energy effective theory would be valid when the canonical kinetic
term $(\partial \Phi)^2$ was large compared to a higher-derivative term such as
\be
  \frac{1}{M^2}(\partial^2 \Phi)^2\, .
\ee
For fluctuations with wavevector $k^\mu = (\omega, \vec k)$, we have $\partial\Phi \sim k \Phi$,
and the lowest-order terms accurately describe the dynamics whenever
$|\vec k| < M$.  A fluctuation that has a low momentum in one frame can, of course,
have a high momentum in some other frame, but the converse is also true; the set of
perturbations that can be safely considered ``low-energy'' looks the same in any frame.

With a Lorentz-violating vector field, the situation is altered.  
In addition to higher-derivative terms of the form $M^{-2}(\partial^2 A)^2$, the
possibility of extra factors of the vector expectation value leads us to consider terms
such as
\be
  {\cal L}_4 = \frac{1}{M^8} A^6 (\partial^2 A)^2\, .
\ee
The number of such higher dimension operators in the effective field
theory is greatly reduced because $A_\m A^\m = -m^2$ and, therefore, 
$A_\m \partial_\n A^\m =0$.  It can be shown that an independent
operator with $n$ derivatives includes at most $2 n$ vector fields, so that the term highlighted
here has the largest number of $A$'s with four derivatives.  We expect that the ultraviolet
cutoff $M$ is of order the vector norm, $M\approx m$.  Hence, when we
consider a background timelike vector field in its rest frame,
\be
  \bar A_\mu = (m, 0, 0, 0)\,,
\ee
the ${\cal L}_4$ term reduces to $m^{-2}(\partial^2 A)^2$, and the effective field theory is
valid for modes with $k< m$, just as in the scalar case.

But now consider a highly boosted frame, with
\be
  \bar A_\mu = (m\cosh\h, m\sinh\h, 0, 0)\,.
\ee
At large $\h$, individual components of $A$ will scale as $e^{|\h|}$, and the
higher-derivative term schematically becomes
\be
  {\cal L}_4 \sim \frac{1}{m^2} e^{6|\h|} (\partial^2 A)^2\,.
\ee
For modes with spatial wave vector $k=|\vec k|$ (as measured in this boosted frame),
we are therefore comparing $m^{-2}e^{6|\h|}k^4$ with the canonical term
$k^2$.  The lowest-order terms therefore only dominate for wave vectors with 
\be
  k < e^{-3|\h|}m\,.
\ee
In the presence of Lorentz violation, therefore, the realm of validity of the effective
field theory may be considerably diminished in highly boosted frames.  We will be
careful in what follows to restrict our conclusions to those that can be reached by
only considering perturbations that are accurately described by the two-derivative terms. 
The instabilities we uncover are infrared phenomena, which cannot be cured by
changing the behavior of the theory in the ultraviolet.
We have been careful to include all of the lowest order terms in the effective field theory expansion---the terms in~\eqref{aLag2}.

\section{Boundedness of the Hamiltonian}

We would like to establish whether there are any values of the parameters $\b_1$,
$\bnew$ and $\b_4$ for which the \aether\  model described above is physically
reasonable.  In practice, we take this to mean that there exist background configurations
that are stable under small perturbations.  It seems hard
to justify taking an unstable background as a starting point for phenomenological 
investigations of experimental constraints, as we would expect the field to evolve
on microscopic timescales away from its starting point.

``Stability'' of a background solution $X_0$ to a set of classical equations of motion 
means that, for any small neighborhood $U_0$ of $X_0$ in the phase space, there is another
neighborhood $U_1$ of $X_0$ such that the time evolution of any point in 
$U_0$ remains in $U_1$ for all times.  More informally, small perturbations oscillate
around the original background, rather than growing with time.  
A standard way of demonstrating stability is to show that the Hamiltonian is a local
minimum at the background under consideration.  Since the Hamiltonian is 
conserved under time evolution, the allowed evolution of a small perturbation will
be bounded to a small neighborhood of that minimum, ensuring stability.
Note that the converse does not necessarily hold; the presence of other conserved
quantities can be enough to ensure stability even if the Hamiltonian is not
bounded from below.  

One might worry about invoking the Hamiltonian in a theory where Lorentz
invariance has been spontaneously violated.  Indeed, as we shall see, the form of
the Hamiltonian for small perturbations will depend on the Lorentz frame in which
they are expressed.  To search for possible linear instabilities, it is necessary to consider
the behavior of small perturbations in every Lorentz frame.

The Hamiltonian density, derived from the action (\ref{vf action}) 
via a Legendre transformation, is
\begin{align} \label{Hamiltonian Density}
{\cal{H}} &= {\partial \lag_A \over \partial(\partial_0 A_\m)} \partial_0 A_\m - \lag_A \\
	&= {\b_1 \over 2} F_{ij}^2 + \b_1 (\partial_0 A_i)^2 -\b_1 (\partial_i A_0)^2 
	+ \bnew(\partial_i A_i)^2  - \bnew (\partial_0 A_0)^2 \nonumber \\
	&\qquad + \beta_4 {A^j A^k \over m^2} (\partial_j A_\rho)(\partial_k A^\rho) 
	- \beta_4 {A^0 A^0 \over m^2} (\partial_0 A_\rho)(\partial_0 A^\rho) ,
\end{align}
where Latin indices $i, j$ run over $\{1, 2, 3\}$. The total Hamiltonian corresponding to this density is
\begin{align}
H &= \int d^3 x \ham \nonumber \\
	&=  \int d^3 x \big( \b_1(\partial_\m A_i \partial_\m A_i - \partial_\m A_0 \partial_\m A_0) 
	+(\b_1- \bnew)[(\pd_0 A_0)^2 - (\pd_i A_i)^2 ]  \nonumber \\ 
	&\qquad+ \beta_4 {A_j A_k \over m^2} (\partial_j A_\rho)(\partial_k A^\rho) 
	- \beta_4 {A_0 A_0 \over m^2} (\partial_0 A_\rho)(\partial_0 A^\rho)\big)\,. 
	\label{hamiltonian}
\end{align} 
We have integrated by parts and assumed that $\partial_i A_j$ vanishes at spatial infinity; repeated lowered indices are summed (without any factors of the metric).  Note that this Hamiltonian is identical to that of a theory with a smooth (positive semi-definite) potential instead of a Lagrange multiplier term, evaluated at field configurations for which the potential is minimized. Therefore, if the Hamiltonian is unbounded when the 
fixed-norm constraint is enforced by a Lagrange multiplier, it will also be unbounded in the
case of a smooth potential.

There are only three dynamical degrees of freedom, so we may reparameterize $A_\m$ such 
that the fixed-norm constraint is automatically enforced and the allowed three-dimensional
subspace is manifest. 
We define a boost variable $\phi$ and angular variables $\theta$ and $\psi$, so that
we can write
\begin{align}
A_0 &\equiv m \cosh \phi \\
A_i &\equiv m \sinh \phi f_i(\theta, \psi) 
\end{align} 
in the timelike case with $A_\m A^\m = - m^2$, and 
\begin{align}
A_0 &\equiv m \sinh \phi \\
A_i &\equiv m \cosh \phi f_i(\theta, \psi) 
\end{align} 
in the spacelike case with $A_\m A^\m = + m^2$.
In these expressions,
\begin{align}
f_1 &\equiv \cos \theta \cos \psi \\
f_2 &\equiv \cos \theta \sin \psi \\
f_3 &\equiv \sin \theta\,,
\end{align}
so that $f_if_i = 1$.
In terms of this parameterization, the Hamiltonian density for a timelike \aether\  field becomes
\begin{align}
\label{tham}
  {\ham^{(t)} \over m^2} &= \b_1\sinh^2\phi \partial_\m f_i \pd_\m f_i  +\b_1\partial_\m \phi \partial_\m \phi
 +(\b_1 - \bnew)\left[ (\pd_0 \phi)^2 \sinh^2\phi - (\cosh \phi f_i \pd_i \phi + \sinh \phi \pd_i f_i)^2 \right] \nonumber \\ 
 &\qquad +\beta_4 \sinh^2\phi \left[ (f_i \partial_i \phi)^2 + \sinh^2\phi (f_i \partial_i f_l)(f_j \partial_j f_l)\right] - \beta_4 \cosh^2\phi \left[ (\partial_0 \phi)^2 + \sinh^2\phi (\partial_0 f_i)^2 \right],
\end{align}
while for the spacelike case we have
\begin{align}
\label{sham}
{\ham^{(s)} \over m^2} &= \b_1 \cosh^2\phi \partial_\m f_i \pd_\m f_i 
  -\b_1 \partial_\m \phi \partial_\m \phi + (\b_1- \bnew) \left[ (\pd_0 \phi)^2 \cosh^2\phi 
 - (\sinh \phi f_i \pd_i \phi + \cosh \phi \pd_i f_i)^2 \right] \nonumber \\
 &\qquad -\beta_4 \cosh^2\phi \left[ (f_i \partial_i \phi)^2 - \cosh^2\phi (f_i \partial_i f_l)(f_j \partial_j f_l)\right] +\beta_4 \sinh^2\phi \left[ (\partial_0 \phi)^2 - \cosh^2\phi (\partial_0 f_i)^2 \right].
\end{align}

Expressed in terms of the variables $\phi, \theta, \psi$, the Hamiltonian is a function of initial data that automatically respects the fixed-norm constraint.  We assume that the derivatives $\partial_\m A_\n (t_0, \vec{x})$ vanish at spatial infinity.

\subsection {Timelike vector field} 

We can now determine which values of the parameters $\{ \b_1, \bnew, \b_4\}$ lead to 
Hamiltonians that are bounded below, starting with the case of a timelike \aether\ 
field.  We can examine the various possible cases in turn.
\begin{itemize}
\item {\bf Case One:  $\b_1=\bnew$ and $\beta_4 = 0$.}

This is the sigma-model kinetic term (\ref{sigmamodel}).
In this case the Hamiltonian density simplifies to
\be
  \ham^{(t)} = m^2 \b_1(\sinh^2\phi \partial_\m f_i \pd_\m f_i  
  +\partial_\m \phi \partial_\m \phi) \,.
\ee
It is manifestly non-negative when $\beta_1 >0$, and non-positive when $\beta_1 < 0$.
The sigma-model choice $\b_1=\bnew >0$ therefore results in a theory that is stable. (See also \S 6.2 of \cite{Eling:2004dk}.)
\item {\bf Case Two: $\b_1 < 0$ and $\beta_4 = 0$.} 

In this case, consider configurations with
 $(\pd_0 f_i) \neq 0$, $(\pd_i f_j) = 0$,  $\pd_\m \phi = 0$, $\sinh^2 \phi \gg 1$.  Then we have
\begin{equation}
{\cal{H}}^{(t)} \sim m^2 \b_1 \sinh^2\phi (\partial_0 f_i)^2.
\end{equation}
For $\b_1<0$, the Hamiltonian can be arbitrarily negative for any value of $\bnew$.

\item {\bf Case Three: $\b_1 \geq 0$,  $\bnew < \b_1$, and $\beta_4 = 0$.}

We consider configurations with 
$\pd_\m f_i = 0$, $f_i \pd_i \phi \neq 0 $, $\pd_0 \phi = 0$,  $\cosh^2 \phi \gg 1 $, which gives
\begin{equation}
{\cal{H}}^{(t)} \sim m^2 (\bnew-\b_1) \cosh^2\phi (f_i \partial_i \phi)^2.
\end{equation}
Again, this can be arbitrarily negative.

\item {\bf Case Four: $\b_1 \geq 0$, $\bnew > \b_1$, and $\beta_4 = 0$.} 

Now we consider configurations with  
$\pd_\m f_i = 0$, $f_i \pd_i \phi = 0$, $\pd_0 \phi \neq 0$, $\sinh^2 \phi \gg 1 $. Then,
\begin{equation}
{\cal{H}}^{(t)} \sim m^2 (\b_1-\bnew) \sinh^2\phi (\partial_0 \phi)^2,
\end{equation}
which can be arbitrarily negative.

\item {\bf Case Five: $\beta_4 \neq 0$.} 

Now we consider configurations with $\pd_\m f_i \neq 0$, $\partial_\m \phi =0$ and $\sinh^2 \phi \gg 1 $. Then,
\begin{equation}
{\cal{H}}^{(t)} \sim m^2 \beta_4 \left[ \sinh^4\phi (f_i \partial_i f_l)(f_k \partial_k f_l) - \sinh^2\phi \cosh^2\phi (\partial_0 f_i)^2\right]\,,
\end{equation}
which can be arbitrarily negative for any non-zero $\beta_4$ and for any values of
$\b_1$ and $\bnew$.  
\end{itemize}
For any case other than the sigma-model choice $\b_1=\bnew$, it is therefore straightforward
to find configurations with arbitrarily negative values of the Hamiltonian.

\begin{figure}
\centering
\includegraphics[width=0.6\textwidth]{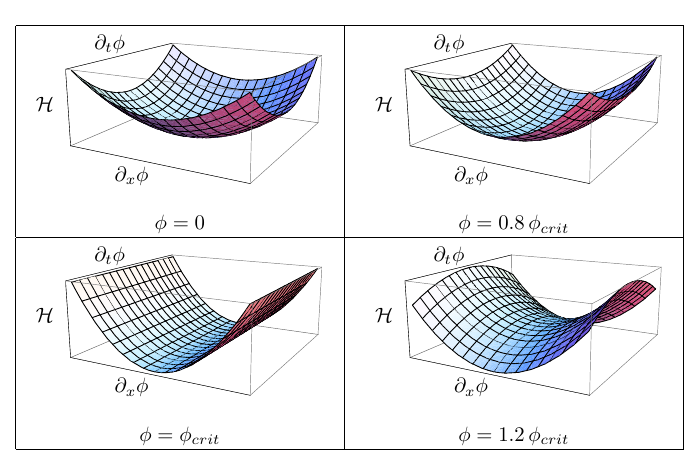}
\caption{Hamiltonian density (vertical axis) when $\b_1 = 1$, $\bnew = 1.1$, and $\theta = \psi = \partial_y \phi = \partial_z \phi = 0$ as a function of $\partial_t \phi$ (axis pointing into page) and $\partial_x \phi$ (axis pointing out of page) for various $\phi$ ranging from zero to $\phi_{crit} = \tanh^{-1} \sqrt{\b_1/\bnew}$, the value of $\phi$ for which the Hamiltonian is flat at $\partial_x \phi = 0$, and beyond. Notice that the Hamiltonian density turns over and becomes negative in the $\partial_t \phi$ direction when $\phi > \phi_{crit}$.}\label{hamiltonian plots}
\end{figure}

Nevertheless, a perturbative analysis of the Hamiltonian would not necessarily discover
that it was unbounded.  The reason for this is shown in
Fig.~\ref {hamiltonian plots}, which shows the Hamiltonian density for the theory
with $\b_1 = 1$, $\bnew = 1.1$,  in a restricted subspace
where $\partial_y\phi = \partial_z\phi = 0$ and $\theta = \phi = 0$, leaving only $\phi$,
$\partial_t\phi$, and $\partial_x\phi$ as independent variables.  We have plotted
$\ham$ as a function of $\partial_t\phi$ and $\partial_x\phi$ for four different values of
$\phi$.
When $\phi$ is sufficiently small, so that the vector is close to being purely timelike,
the point $\partial_t\phi = \partial_x\phi = 0 $ is a local minimum.  Consequently,
perturbations about constant configurations with small $\phi$ would appear stable.  But for large values of $\phi$, the unboundedness of the Hamiltonian
becomes apparent.  This phenomenon will arise again when we consider the evolution
of small perturbations in the next section. At the end of this section, we will explain why such regions of large $\phi$ are still in the regime of validity of the effective field theory expansion.

\subsection{Spacelike vector field}

We now perform an equivalent analysis for an \aether\ field with a spacelike expectation
value.  In this case all of the possibilities lead to Hamiltonians \eqref{sham}  that are unbounded below,
and the case $\b_1=\bnew > 0$ is not picked out.

\begin{itemize}

\item {\bf Case  One:  $\b_1 < 0$ and $\beta_4 = 0$.} 

Taking  $(\pd_\m \phi)  = 0$,  $\pd_j f_i = 0$, $\pd_0 f_ i \neq 0$, we find
\begin{equation}
{\cal{H}}^{(s)} \sim m^2 \b_1\cosh^2\phi (\partial_0 f_i)^2.
\end{equation}

\item {\bf Case Two: $\b_1 > 0$, $\bnew \leq \b_1$, and $\beta_4 = 0$.}    

Now we consider $\pd_\m f_i = 0$, $\pd_i \phi \neq 0 $, $\pd_0 \phi = 0$,  giving
\begin{equation}
{\cal{H}}^{(s)} \sim m^2 \left[ - \b_1 \pd_i \phi \pd_i \phi + (\bnew-\b_1) \sinh^2\phi (f_i \partial_i \phi)^2\right].
\end{equation}

\item {\bf Case  Three: $\b_1 \geq 0$, $\bnew > \b_1$, and $\beta_4 = 0$.}   

In this case we examine $(\pd_0 \phi) \neq 0$,  $\pd_\m f_i = 0$, $\pd_i \phi = 0$, which leads to
\begin{equation}
{\cal{H}}^{(s)} \sim m^2 (\b_1-\bnew) \cosh^2\phi (\partial_0 \phi)^2.
\end{equation}

\item {\bf Case Four: $\beta_4 \neq 0$.}   

Now we consider configurations with $\pd_\m f_i \neq 0$, $\partial_\m \phi =0$ and $\sinh^2 \phi \gg 1 $. Then,
\begin{equation}
{\cal{H}}^{(s)} \sim m^2 \beta_4 \left( \cosh^4\phi (f_i \partial_i f_l)(f_k \partial_k f_l) - \cosh^2\phi \sinh^2\phi (\partial_0 f_i)^2\right).
\end{equation}
\end{itemize}
In every case, it is clear that we can find initial data for a spacelike vector field that makes the
Hamiltonian as negative as we please, for all possible $\b_1$, $\beta_4$ and $\bnew$.


\subsection{Smooth Potential}

The usual interpretation of a Lagrange multiplier constraint is that it is the low-energy limit of smooth potentials when the massive degrees of freedom associated with excitations away from the minimum cannot be excited.  We now investigate whether these degrees of freedom can destabilize the theory.  Consider the most general, dimension four, positive semi-definite smooth potential that has a minimum when the vector field takes a timelike vacuum expectation value,
\begin{equation}
V = {\xi \over 4} (A_\m A^\m + m^2)^2,
\end{equation}
where $\xi$ is a positive dimensionless parameter.  The precise form of the potential should not affect the results as long as the potential is non-negative and has the global minimum at $A_\m A^\m = -m^2$.

We have seen that the Hamiltonian is unbounded from below unless the kinetic term takes the sigma-model form, $(\partial_\m A_\n)(\partial^\m A^\n)$.  Thus we take the Lagrangian to be
\begin{equation}
{\cal{L}} = -{1\over 2}(\partial_\m A_\n)(\partial^\m A^\n) - {\xi \over 4} (A_\m A^\m + m^2)^2.
\end{equation}

Consider some fixed timelike vacuum $\bar A_\mu$ satisfying $\bar{A}_\m \bar{A}^\m = -m^2$.  
We may decompose the \aether\ field into a scaling of the norm, represented by a scalar $\Phi$,
and an orthogonal displacement, represented by vector $B_\mu$ satisfying $\bar{A}_\mu B^\mu = 0$.
We thus have
\be
  A_\m = \bar{A}_\m - {\bar{A}_\m \Phi \over m} + B_\m\,,
\ee
where
\begin{equation}
B_\m = \left(\eta_{\m\n}+{\bar{A}_\m\bar{A}_\n \over m^2}\right) A^\n~~\text{and}~~~ \Phi = {\bar{A}_\m A^\m \over m}+m.
\end{equation}
With this parameterization, the Lagrangian is
\begin{equation}
{\cal{L}} = {1 \over 2} (\partial_\m \Phi)( \partial^\m \Phi) - {1 \over 2} (\partial_\m B_\n)(\partial^\m B^\n) -{\xi \over 4} (2m\Phi + B_\m B^\m - \Phi^2)^2.
\end{equation}
The field $\Phi$ automatically has a wrong sign kinetic term, and, at the linear level, propagates with a dispersion relation of the form
\begin{equation}
\omega_\Phi^2 = \vec{k}^2 - 2\xi m^2.
\end{equation}
We see that in the case of a smooth potential, there exists a ghostlike mode (wrong-sign kinetic term) that is also tachyonic with spacelike wave vector and a group velocity that generically exceeds the speed of light.   It is easy to see that sufficiently long-wavelength perturbations will exhibit exponential growth.  The existence of a ghost when the norm of the vector field is not strictly fixed was shown in \cite{Elliott:2005va}.   

In the limit as $\xi$ goes to infinity, the equations of motion enforce a fixed-norm constraint and the ghostlike and tachyonic degree of freedom freezes. The theory is equivalent to one of a Lagrange multiplier if the limit is taken appropriately. 

\subsection{Discussion}

To summarize, we have found that the action in~\eqref{vf action} leads to a Hamiltonian that is globally bounded from below only in the case of a timelike sigma-model Lagrangian,
corresponding to $\b_1 = \bnew > 0$ and $\beta_4 = 0$.  Furthermore, we have verified (as was shown in \cite{Elliott:2005va}) that if the Lagrange multiplier term is replaced by a smooth, positive semi-definite potential, then a tachyonic ghost propagates and the theory is destabilized. 

If the Hamiltonian is bounded below, the theory is stable, but the converse is not
necessarily true.  The sigma-model theory is the only one for which this criterion suffices to guarantee
stability.  In the next section, we will examine the linear stability of these models by
considering the growth of perturbations.  Although some models are stable at the linear
level, we will see in the following section that most of these have negative-energy ghosts,
and are therefore unstable once interactions are included.  The only exceptions, both
ghost-free and linearly stable, are the
Maxwell \eqref{maxwell} and scalar \eqref{scalar} 
models.  

We showed in the previous section that, unless 
$\bnew - \beta_1$ and $\beta_4$ are exactly zero, the
Hamiltonian is unbounded from below.
However, the effective field theory breaks down before arbitrarily
negative values of the Hamiltonian can be reached; when $\bnew \neq
\b_1$ and/or $\b_4 \neq 0$, in regions of phase space in which ${\cal H} < 0$ (schematically),
\be
{\cal H} \sim - m^2 e^{4 |\phi|} (\pd \Theta)^2 \qquad \text{where} \qquad \Theta \in \{\phi, \theta, \psi\}.
\ee The effective field theory breaks down when kinetic terms with four derivatives (the terms of next highest order in the effective field theory expansion) are on the order of terms with two derivatives, or, in the angle parameterization, when
\be
m^2 e^{4 |\phi|} (\pd \Theta)^2 \sim  e^{8 |\phi|} (\pd \Theta)^4.
\ee
In other words, the effective field theory is only valid when
\be
e^{2 |\phi|} |\pd \Theta | < m.
\ee 
In principle, terms in the effective action with four or more derivatives could 
add positive contributions to the Hamiltonian to make it bounded from below.
However, our analysis shows that the Hamiltonian (in models other than the timelike sigma model
with fixed norm) is necessarily concave down around the 
set of configurations with constant \aether\ fields.  If higher-derivative terms intervene to
stabilize the Hamiltonian, the true vacuum would not have $H=0$.
Theories could also be deemed stable if there are additional symmetries that lead to conserved currents (other than energy-momentum density) or to a reduced number of physical degrees of freedom. 

Regardless of the presence of terms beyond leading order in the
effective field theory expansion, due to the presence of the
ghost-like and tachyonic mode (found in the previous section),
there is an unavoidable problem with
perturbations when the field moves in a smooth, positive semi-definite
potential.   This exponential instability will be present regardless
of higher order terms in the effective field theory expansion because it occurs for very long-wavelength modes
(at least around constant-field backgrounds).

\section{Linear instabilities} 

We have found that the Hamiltonian of a generic \aether\ model is unbounded below.
In this section, we investigate whether there exist actual physical instabilities at the
linear level---{\it i.e.}, whether small perturbations grow exponentially with time.
It will be necessary to consider the behavior of small fluctuations in every 
Lorentz frame,\footnote{The theory of perturbations about a constant background is 
equivalent to a theory with explicit Lorentz violation because the first order
Lagrange density includes the term, $\lambda \bA^\m \d A_\m$, where
$\bA^\m$ is effectively some constant coefficient.}  not
only in the \aether\ rest frame  \cite{Kostelecky:2001xz, Mattingly:2005re,Adams:2006sv}.
We find a range of parameters $\beta_i$ for which the theories are tachyon-free; these
correspond (unsurprisingly) to dispersion relations for which the phase velocity
satisfies $0 \leq v^2 \leq 1$.  In \S \ref{ghosts} we consider the existence of ghosts.

\subsection{Timelike vector field}

Suppose Lorentz invariance is spontaneously broken so that there is a preferred rest frame, and imagine that perturbations of some field in that frame have the following dispersion relation:
\begin{equation} \label{disprelation}
v^{-2} \o^2 = \kv \cdot \kv.
\end{equation}
This can be written in frame-invariant notation as
\begin{equation}\label{Tdispersion}
(v^{-2} - 1) (t^\m k_\m)^2 = k_\m k^\m,
\end{equation} where $t^\m$ is a timelike Lorentz vector that characterizes the
4-velocity of the preferred rest frame. So, in the rest frame, $t^\m =
\{1,0,0,0\}$. Indeed, in the Appendix, we find dispersion relations for the \aether\ modes of exactly the form in \eqref{Tdispersion} with $t^\m = \bar{A}^\m / m$ and \eqref {apA:spin-1}
\begin{equation}
v^2 = {\b_1 \over \b_1 - \b_4} 
\end{equation} and \eqref{scalar mode}
\begin{equation}
v^2 = {\b_* \over \b_1 - \b_4}.
\end{equation}

Now consider the dispersion relation for perturbations
of the field in another (``primed'') frame. Let's solve for $k_0' =
\o'$, the frequency of perturbations in the new 
frame. Expanded out, the dispersion relation reads
\begin{equation}
\o'^2(1 + (v^{-2}-1)(t'^0)^2) + 2 \o' (v^{-2} - 1)t'^0 t'^i k_i' 
- \kv' \cdot \kv' + (v^{-2} - 1) (t'^i k_i')^2 = 0
\end{equation}
where $i \in \{ 1,2,3 \}$. The solution for $\o'$ is:
\begin{equation}
\label{omegaTdispersion}
\o' = {-(v^{-2} - 1)t'^0 t'^i k'_i  \pm \sqrt{D_{(t)}} \over 1 +(v^{-2}-1)(t'^0)^2}\,,
\end{equation} 
where
\begin{equation}
{D_{(t)}} = \kv' \cdot \kv' + (v^{-2}-1)\left((t'^0)^2 \kv' \cdot \kv' - (t'^i k'_i)^2 \right).
\end{equation}
In general, $t'^0 = \cosh \h$ and $t'^i = \sinh \h \, \hat{n}^i$, where $\hat{n}_i \hat{n}^i
= 1$ and $\h = \cosh^{-1}\g$ is a boost parameter. We therefore have
\begin{equation}
{D_{(t)}} = \kv' \cdot \kv' \left\{1 + (v^{-2}-1)\left[\cosh^2\h  -
  \sinh^2\h \, (\hat{n}\cdot  \hat{k}')^2 \right] \right\},
\end{equation} 
where $\hat{k}' = \vec{k}'/|\vec{k}'|$.
Thus ${D_{(t)}} $ is clearly greater than zero if $v \leq 1$. However, if $v > 1$ then ${D_{(t)}} $ can be negative for very large boosts if $\kv'$ is not parallel to the boost direction. 

The sign of the discriminant ${D_{(t)}} $ determines whether the frequency
$\o'$ is real- or complex-valued. We have shown that when the phase
velocity $v$ of some field excitation is greater than the speed of
light in a preferred rest frame, then there is a (highly boosted)
frame in which the excitation looks unstable---that is, the frequency
of the field excitation can be imaginary. More specifically, plane
waves traveling along the boost direction with boost parameter $\gamma = \cosh \h$ have a growing
amplitude if $\gamma^2 > 1/(1-v^{-2}) > 0$.

In Appendix~\ref{Ap:A}, we find dispersion relations of the form in~\eqref{Tdispersion} for the various massless excitations about a constant timelike background ($t^\m = \bA^\m /m$). Requiring stability and thus $0 \leq v^2 \leq 1$ leads to the inequalities,
\begin{equation} \label{spin1timelike}
0 \le {\beta_1 \over \beta_1 - \beta_4} \le 1
\end{equation}
and
\begin{equation} \label{spin0timelike}
0 \le {\bnew \over \beta_1 - \beta_4} \le 1 \, .
\end{equation}
Models satisfying these relations are stable with respect to linear perturbations in any
Lorentz frame.

\subsection{Spacelike vector field}\label{s superluminal}

We show in Appendix~\ref{Ap:A} that fluctuations about a spacelike,
fixed-norm, vector field background have dispersion relations of the form
\begin{equation}\label{Sdispersion}
(v^2 - 1) (s^\m k_\m)^2 = - k_\m k^\m,
\end{equation} with $s^\m = \bar{A}^\m / m$ and \eqref {apA:spin-1}
\begin{equation}
v^2 = { \b_1 + \b_4 \over \b_1} 
\end{equation} and \eqref{scalar mode}
\begin{equation}
v^2 = {  \b_1 + \b_4 \over \b_*}.
\end{equation}
 In frames where $s^\m = \{0, \hat{s}\}$, 
$v$ is the phase velocity in the $\hat{s}$ direction.  

Consider solving for $k'_0 = \omega'$ in an arbitrary (``primed'') frame. The
solution is as in~\eqref{omegaTdispersion}, but with $v^{-2} \rightarrow
2 - v^{2}$ and $t'^\m \rightarrow s'^\m$. Thus, 
\begin{equation}
\label{omegaSdispersion}
\o' = {(v^{2} - 1)s'^0 s'^i k'_i  \pm \sqrt{D_{(s)}} \over 1 +(1-
v^{2})(s'^0)^2}\,,
\end{equation} 
where
\begin{equation}
D_{(s)} = \kv' \cdot \kv' - (v^{2}-1)\left[(s'^0)^2 \kv' \cdot \kv' - (s'^i k'_i)^2 \right].
\end{equation}
In general, $s'^0 = \sinh \h$ and $s'^i = \cosh \h \, \hat{n}^i$ where $\hat{n}_i \hat{n}^i
= 1$ and $\h = \cosh^{-1}\g$ is a boost parameter. So,
\begin{equation}
D_{(s)} = \kv' \cdot \kv' \left\{ 1 - (v^{2}-1)\left[\sinh^2\h  -
  \cosh^2\h \, (\hat{n}\cdot  \hat{k'})^2 \right] \right\}.
\end{equation}
which can be rewritten,
\begin{equation}
D_{(s)} = \kv' \cdot \kv'\left\{ v^{2} + (1-v^{2})\cosh^2\h\left[1-
    (\hat{n}\cdot \hat{k'})^2 \right]\right\}.
\end{equation}
It is clear that $D_{(s)}$ is non-negative for all values of $\eta$ if and only if $ 0 \leq v^2 \leq 1$. 
The theory will be unstable unless $0 \leq v^2 \leq 1$.

The dispersion relations of the form~\eqref{Sdispersion} for the massless excitations about the spacelike background are given in Appendix~\ref{Ap:A}.   The requirement that $ 0 \leq v^2 \leq 1$ implies
\begin{equation} \label{spin1spacelike}
0 \le {\beta_1 + \beta_4 \over \beta_1} \le 1 
\end{equation}
and
\begin{equation} \label{spin0spacelike}
0 \le {\beta_1 + \beta_4 \over \bnew} \le 1 \, .
\end{equation}
Models of spacelike \aether\ fields will only be stable with respect to linear perturbations
if these relations are statisfied.

The requirements \eqref{spin0timelike} or \eqref{spin0spacelike} do not apply in the Maxwell case
(when $\b_*=0=\b_4$), and those of \eqref{spin1timelike} or \eqref{spin1spacelike} do not 
apply in the scalar case (when $\b_1 = 0 = \b_4$), since the corresponding degrees of freedom in each case do not propagate. 

\subsection{Stability is not frame-dependent}

The excitations about a constant background are massless (\emph{i.e.}~the frequency is proportional to the magnitude of the spatial wave vector), but they generally do not propagate along the light cone. In fact, when $v >1$, the wave vector is timelike even though the cone along which excitations propagate is strictly outside the light cone.  We have shown that such excitations blow up in some frame. 
The exponential instability occurs for observers in boosted frames. In these frames, portions of constant-time hypersurfaces are actually inside the cone along which excitations propagate.

Why do we see the instability in only \emph{some} frames when performing a linear stability analysis? 
Consider boosting the wave four-vectors of such excitations with
complex-valued frequencies and real-valued spatial wave vectors back
to the rest frame. Then, in the rest frame, both the frequency and the
spatial wave vector will have non-zero imaginary parts. Such solutions
with complex-valued $\vec{k}$ require initial data that grow at
spatial infinity and are therefore not really ``perturbations'' of the
background.  But even though the \aether\ field defines a rest frame,
there is no restriction against considering small perturbations
defined on a constant-time hypersurface in any frame.  Well-behaved
initial data can be decomposed into modes with real spatial wave
vectors; if any such modes lead to runaway growth, the theory is
unstable.

\section{Negative Energy Modes} \label{ghosts}

We found above that manifest perturbative stability in all frames requires $0\le v^2 \le 1$. In the Appendix, we show that there are two kinds of propagating modes, except when $\b_*= \b_4 = 0$ or when $\b_1 = \b_4 = 0$. Based on the dispersion relations for these modes, the $0\le v^2 \le 1$ stability requirements translated into the inequalities for $\b_*, \b_1$, and $\b_4$ in~\eqref{spin1timelike}-\eqref{spin0timelike} for timelike \aether\ and~\eqref{spin1spacelike}-\eqref{spin0spacelike} for spacelike {\ae}ther. We shall henceforth assume that these inequalities hold and, therefore, that $\o$ and $\vec{k}$ for each mode are real in every frame.  We will now show that, even when these requirements
are satisfied and the theories are linearly stable, there will be negative-energy ghosts that imply
instabilities at the nonlinear level (except for the sigma model, Maxwell, and scalar cases).

For timelike vector fields, with respect to the {\ae}ther rest frame, the various modes correspond to two spin-1 degrees of freedom and one spin-0 degree of freedom. Based on their similarity in form to the timelike {\ae}ther rest frame modes, we will label these modes once and for all as ``spin-1'' or ``spin-0,'' even though these classifications are only technically correct for timelike fields in the {\ae}ther rest frame. 

The solutions to the first order equations of motion for perturbations
$\d A_\m$ about an arbitrary, constant, background $\bA_\m$ satisfying $\bA^\m \bA_\m \pm m^2 = 0$
are (see Appendix~\ref{Ap:A}):
\begin{equation}
\d A_\m = \int d^4 k \, q_\m(k) e^{i k_\m x^\m}, \qquad q_\m(k) = q_\m^*(-k)
\end{equation} where either,
\begin{equation}
\label{mode1}
q_\m(k) = i \a^\n k^\r {\bA^\s\over m} \e_{\m \n \r
  \s}~~\text{and}~~\b_1 k_\m k^\m + \b_4{ ( \bA_\m k^\m)^2 \over m^2}= 0~~\text{and}~~\a^\n \bA_\n = 0   \qquad \text{(spin-1)}
\end{equation} where $\a^\n$ are real-valued constants or
\begin{equation}
\label{mode2}
q_\m =  i \a \left(\h_{\m \n} \pm {\bA_{\m} \bA_{\n}
    \over m^2}\right)k^\n \qquad 
\text{and} \qquad \left( \bnew \h_{\m \n}  + \left(\b_4 \pm (\bnew -
  \b_1) \right) {\bA_{\m} \bA_{\n}\over m^2}\right)k^\m k^\n = 0  \qquad \text{(spin-0)}
\end{equation}
where $\a$ is a real-valued constant. 

Note that when $\b_1 = \b_4 = 0$, corresponding to the scalar form of \eqref{scalar}, the spin-1 dispersion relation is satisfied  trivially, because the spin-1 mode does not propagate in this case. Similarly, when $\b_* = \b_4 = 0$, the kinetic term takes on the Maxwell form in \eqref{maxwell} and the spin-0 dispersion relation becomes $\bA_\m k^\m = 0$; the spin-0 mode does not propagate in that case.

The Hamiltonian~\eqref{hamiltonian} for either of these modes is
\begin{equation}
\label{k hamiltonian}
	H = \int d^3 k \, \left\{ \left[ \b_1 (\o^2+\kv\cdot \kv) +\b_4(-(\ba^0 \o)^2 + (\ba^i k_i)^2) \right]  q^\m q_\m^{*} + (\b_1 - \bnew) (\o^2 q_0^*q_0 + k_i q_i^* k_j q_j ) \right\}\,,
\end{equation}
where $k_0 = \o = \o(\kv)$ is given by the solution to a dispersion relation and where 
$\ba^\m \equiv \bA^\m / m$.  
One can show that, as long as $\b_1$ and $\b_4$ satisfy the conditions~\eqref{spin1timelike} 
or~\eqref{spin1spacelike} that guarantee real frequencies $\o$ in all frames, we will have
\be
  q^*_\m q^\m \geq 0
\ee 
for all timelike and spacelike vector perturbations.
We will now proceed to evaluate the Hamiltonian for each mode in different
theories.

\subsection{Spin-1 energies}

In this section we consider nonvanishing $\b_4$, and show that the spin-1 mode can carry
negative energy even when the conditions for linear stability are satisfied.

\paragraph*{Timelike vector field.}
Without loss of generality, set 
\be
\bA_\m = m (\cosh \h, \sinh \h \, \hat{n}).
\ee where $\hat{n} \cdot \hat{n} = 1$.
The energy of the spin-1 mode in the timelike case is given by
\begin{equation}
H =  \int d^3k (\vec{k}\cdot \vec{k}) q^*_\m q^\m \left[{2 X \mp \beta_4 \sinh(2\h) (\hat{n}\cdot\hat{k})\sqrt{X} \over \beta_1 -\beta_4 \cosh^2\h } \right],
\end{equation}
where
\begin{equation}
X = \beta_1\left\{\beta_1 + \beta_4\left[(\hat{n}\cdot \hat{k})^2 \sinh^2\h - \cosh^2\h
\right]\right\}.  \end{equation}
Looking specifically at modes for which $\hat{n} \cdot \hat{k} = +1$, we find
\begin{equation}
H =  \int d^3k (\vec{k}\cdot \vec{k})q^*_\m q^\m \left[{2\beta_1(\beta_1 -  \beta_4) \mp \beta_4 \sinh(2\h) \sqrt{\beta_1(\beta_1 -  \beta_4)} \over \beta_1 -\beta_4 \cosh^2\h } \right]\,.
\end{equation}
The energy of such a spin-1 perturbation can be negative when
$|\beta_4 \sinh(2\h)| > 2\sqrt{\beta_1(\beta_1 -  \beta_4)}$.  Thus it is possible to have negative energy perturbations whenever $\beta_4 \neq 0$.   Perturbations with wave numbers perpendicular to the boost direction have positive semi-definite energies.

\paragraph*{Spacelike vector field.}
Without loss of generality, for the spacelike case we set
\be
\bA_\m = m (\sinh \h, \cosh \h\,\hat{n})\,,
\ee where $\hat{n} \cdot \hat{n} = 1$.
The energy of the spin-1 mode in this case is given by
\begin{equation}
H =  \int d^3k (\vec{k}\cdot \vec{k}) q^*_\m q^\m
\left[{2 X \mp \beta_4 \sinh(2\h) (\hat{n}\cdot\hat{k})\sqrt{X} \over \beta_1 -\beta_4 \sinh^2\h } \right],
\end{equation}
where
\begin{equation}
X = \beta_1\left\{\beta_1 + \beta_4\left[(\hat{n}\cdot \hat{k})^2 \cosh^2\h - \sinh^2\h\right]\right\}.  
\end{equation}
Looking at modes for which $\hat{n} \cdot \hat{k} = +1$, we find
\begin{equation}
H =  \int d^3k (\vec{k}\cdot \vec{k})q^*_\m q^\m \left[{2\beta_1(\beta_1 +  \beta_4) \mp \beta_4 \sinh(2\h) \sqrt{\beta_1(\beta_1 +  \beta_4)} \over \beta_1 - \beta_4 \sinh^2\h } \right]\,.
\end{equation}
Thus, the energy of perturbations can be negative when $|\beta_4 \sinh(2\h)| > 2\sqrt{\beta_1(\beta_1 +  \beta_4)}$.  Thus it is possible to have negative energy perturbations whenever $\beta_4 \neq 0$.  Perturbations with wave numbers perpendicular to the boost direction have positive semi-definite energies. In either the timelike or spacelike case, models with $\b_4\neq 0$ feature spin-1
modes that can be ghostlike.

We note that the effective field theory is valid when $k < e^{- 3 |\h|} m$, as detailed in \S \ref{Validity of effective field theory}. But even if $\h$ is very large, the effective field theory is still valid for very long wavelength perturbations, and therefore such long wavelength modes with negative energies lead to genuine instabilities.

\subsection{Spin-0 energies}

We now assume the inequalities required for linear stability, \eqref{spin0timelike} or~\eqref{spin0spacelike}, and also that $\b_4 = 0$. We showed above that, otherwise, there are growing modes in some frame or there are propagating spin-1 modes that have negative energy in some frame. 
When $\bnew \neq 0$, the energy of the spin-0 mode in~\eqref{mode2} is given by
\begin{equation}
\label{mode2 energy}
	H = 2 \b_1 \a^2 \int d^3 k \, (\ba_\r k^\r)^2\left( \o^2(\kv)\left[\pm 1 - (1 - \b_1/\bnew) \ba_0^2 \right] + \o(\kv)\, \ba_0 (1 - \b_1/\bnew) \ba_i k_i \right)
\end{equation} for $\bA_\m \bA^\m \pm m^2 = 0$ and $\ba_\m \equiv \bA_\m / m$.   

\paragraph*{Timelike vector field.}
We will now show that the quadratic order Hamiltonian can be negative when the background is timelike and the kinetic term does not take one of the special forms (sigma model, Maxwell, or scalar).
Without loss of generality we set $\ba_0 = \cosh \h $ and $\ba_i = \sinh \h \, \hat{n}_i$, where $\hat{n}\cdot \hat{n} = 1$.  Then plugging the freqency $\o(\vec{k})$, as defined by the spin-0 dispersion relation, into the Hamiltonian~\eqref{mode2 energy} gives 
\begin{equation}
\label{T spin-0 energy}
	H = \b_1 \a^2 \int d^3 k \, (\ba_\r k^\r)^2 \left[{
	2 X \pm (1- \beta_1/\bnew)\sinh 2\h  (\hat{n}\cdot \hat{k})\sqrt{X}
	\over
	1 + (\b_1/\bnew - 1)\cosh^2\h
	}\right],
\end{equation} 
where
\begin{equation}
X = {1+ (\b_1/\bnew -1) [\cosh^2\h - (\hat{n}\cdot\hat{k})^2\sinh^2\h ]}.
\end{equation}
If $\hat{n}\cdot\hat{k} \neq 0$, the energy can be negative. In particular, if $\hat{n}\cdot\hat{k} = 1$ we have 
\begin{equation}
H = \b_1 \a^2 \int d^3 k \, (\ba_\r k^\r)^2 \left[{
	2 \b_1/\bnew \pm (1- \beta_1/\bnew)\sinh 2\h \sqrt{\b_1 / \bnew}
	\over
	1+ (\b_1/\bnew - 1)\cosh^2\h
	}\right].
\end{equation} 
Given that $ \b_1 / \bnew -1 \geq 0$, $H$ can be  negative when $| \sinh 2 \h | > 2 \sqrt{\b_1 / \bnew} / (\b_1 / \bnew - 1)$.  

We have thus shown that, for timelike backgrounds, there are modes that in some frame have 
negative energies and/or growing amplitudes as long as $\b_1 \neq \bnew$, $\b_1 \neq 0$, 
and $\bnew\neq 0$. Therefore, the only possibly stable theories of timelike \aether\ fields
are the special cases mentioned earlier:  the sigma-model ($\b_1 = \bnew$), Maxwell 
($\bnew = 0$), and scalar ($\b_1=0$) kinetic terms.

\paragraph*{Spacelike vector field.}  For the spacelike case,
without loss of generality we set $\ba_0 = \sinh \h $ and $\ba_i = \cosh \h \, \hat{n}_i$, 
where $\hat{n}\cdot \hat{n} = 1$.  Once again, plugging
the frequency $\o(k)$ into the Hamiltonian~\eqref{mode2 energy} gives
\begin{equation}
\label{S spin-0 energy}
	H = \b_1 \a^2 \int d^3 k \, (\ba_\r k^\r)^2 \left[{
	- 2 X \pm (1 - \beta_1/\beta_*)\sinh 2\h  (\hat{n}\cdot \hat{k})\sqrt{X}
	\over
	1 + (1- \b_1 / \bnew)\sinh^2\h
	}\right] ,
\end{equation} 
where
\begin{equation}
X = {1 + (1- \b_1 / \bnew ) \left[\sinh^2\h - (\hat{n}\cdot\hat{k})^2\cosh^2\h \right]}.
\end{equation}
Upon inspection, one can see that there are values of 
$\hat{n}\cdot \hat{k}$ and $\h$ that make $H$ negative,  except when $\bnew = 0$ 
(Maxwell) or $\b_1 = 0$ (scalar).
Again, the Hamiltonian density is less than zero for modes with wavelengths sufficiently long 
($k < e^{-3 |\h|} m$), so the effective theory is valid.

\section{Maxwell and Scalar Theories}

We have shown that the only version of the \aether\ theory \eqref{vf action} for which the
Hamiltonian is bounded below is the timelike
sigma-model theory ${\cal L}_K = -(1/2)(\partial_\mu A_\nu)
(\partial^\mu A^\nu)$, corresponding to the choices $\b_1=\bnew$, $\b_4=0$, with the
fixed-norm condition imposed by a Lagrange multiplier constraint.  (Here and below, we
rescale the field to canonically normalize the kinetic terms.)
However, when we looked for explicit instabilities in the form of
tachyons or ghosts in the last two sections, we found two other models for which such
pathologies are absent:  the Maxwell Lagrangian
\be
  {\cal L}_K = -\frac{1}{4}F_{\mu\nu}F^{\mu\nu}\, ,
  \label{maxwell2}
\ee
corresponding to $\bnew = 0 = \b_4$, and the scalar Lagrangian
\be
  {\cal L}_K = \frac{1}{2} (\partial_\mu A^\mu)^2\, ,
\ee
corresponding to $\b_1 = 0 = \b_4$.
In both of these cases, we found that the Hamiltonian is unbounded below,\footnote{Boundedness of the Hamiltonian was considered in \cite{Clayton:2001vy}.} but 
a configuration with a small positive energy does not appear to run away into an
unbounded region of phase space characterized by large negative and positive balancing
contributions to the total energy.  

These two models are also distinguished in another way:
there are fewer than three propagating degrees of freedom at first order in perturbations in the Maxwell and scalar Lagrangian cases, while there are three in all others.  This is closely tied
to the absence of perturbative instabilities;  the ultimate cause of those instabilities can
be traced to the difficulty in making all of the degrees of freedom simultaneously well-behaved.
The drop in number of degrees of freedom stems from the fact that $A_0$ lacks time derivatives in the Maxwell Lagrangian and that the $A_i$ lack time derivatives in the scalar Lagrangian. In other words, some of the vector components are themselves Lagrange multipliers in these special cases. 

Only two perturbative degrees of freedom---the spin-1 modes---propagate in the Maxwell case (cf.~\eqref{mode1}-\eqref{mode2} when $\b_* = 0 = \b_4$). The ``mode'' in~\eqref{mode2} is a gauge degree of freedom; at first order in perturbations the Lagrangian has a gauge-like symmetry under $\d A_\m \rightarrow \d A_\m + \partial_\m \phi(x)$ where $\bA^\m \partial_\m \phi = 0$.  As expected of a gauge degree of freedom, the spin-0 mode has zero energy and does not propagate. Meanwhile, the spin-1 perturbations propagate as well-behaved plane waves and have positive energy.  We note that the Dirac method for counting degrees of freedom in constrained dynamical systems implies that there are \emph{three} degrees of freedom \cite{Bluhm:2008yt}.\footnote{For a discussion of constrained dynamical systems see \cite{Henneaux:1992ir}.} The additional degree of freedom, not apparent at the linear level, could conceivably cause an instability; this mode does not propagate because it is gauge-like at the linear level, but there is no gauge symmetry in the full theory.

In the scalar case, there are no propagating spin-1 degrees of freedom.  The spin-0 degree of freedom has a nontrivial dispersion relation but no energy density (cf.~\eqref{mode1}-\eqref{mode2},~\eqref {T spin-0 energy}, and \eqref {S spin-0 energy} when $\b_1 = 0 = \b_4$) at leading order in the
perturbations.  Essentially, the fixed-norm constraint is incompatible with what would be
a single propagating scalar mode in this model; the theory is still dynamical, but perturbation
theory fails to capture its dynamical content.

Each of these models displays some idiosyncratic features, which we now consider in turn.

\subsection{Maxwell action}

The equation of motion for the Maxwell Lagrangian with a fixed-norm constraint is
\be
  \partial_\mu F^{\mu\nu} = -2\lambda A^\nu\,.
\ee
Setting $A_\mu A^\mu = \mp m^2$, the Lagrange multiplier is given by
\be
  \lambda = \pm \frac{1}{2m^2}A_\nu \partial_\mu F^{\mu\nu}\,.
\ee
For timelike {\ae}ther fields, the sign of $\l$ is preserved along timelike trajectories since, when the kinetic term takes the special Maxwell form, there is a conserved current (in addition to energy-momentum density) due to the Bianchi identity\footnote{If $\l > 0$ initially, then it must pass through $\l =0$ to reach $\l < 0$---but $\l = 0$ is conserved along timelike trajectories, so $\l$ can at best stop at $\l = 0$.}:
\begin{equation}
  \label{eq:conservedcurrent}
 0 = \partial_\n  (\partial_\m F^{\m \n}) = -2 \partial_\n (\lambda A^\n).
\end{equation}
In particular, the condition that $\lambda = 0$ is conserved along timelike $A^\n$ 
\cite{Jacobson:2000xp,Bluhm:2008yt}.  In the presence of interactions this will continue to be true 
only if the coupling to external sources takes the form of an
interaction with a conserved current, $A_\mu J^\mu$ with $\partial_\mu J^\mu=0$.

If we take the timelike Maxwell theory coupled to a conserved current
and restrict to initial data satisfying $\lambda = 0$
at every point in space, the theory reduces precisely to Maxwell electrodynamics---not only in the equation of motion, but also in the energy-momentum tensor.
We can therefore be confident that this theory, restricted to this subset of initial
data, is perfectly well-behaved, simply because it is identical to conventional electromagnetism
in a nonlinear gauge \cite{Nambu:1968qk, Chkareuli:2006yf, Bluhm:2007bd}. 

In the case of a spacelike vector expectation value, there is an explicit obstruction to
finding smooth time evolution for generic initial data.  In this case, the constraint equations are
\begin{equation}
- A_0^2 + A_i A_i  = m^2 \qquad \text{and} \qquad \pd_i \pd^i A_0 - \pd_0 \pd_i A^i = -2\l A_0.
\end{equation} 
Suppose spatially homogeneous initial conditions for the $A_i$ are given. Without loss of generality, we can align axes such that
\begin{equation}
A_\mu(t_0) = (A_0(t_0),0,0,A_3(t_0)),
\end{equation}
where $-A^2_0 + A^2_3 = m^2$. If $A_i A_i \neq m^2$, the equations of motion are 
\begin{equation}\label{badexample}
\pd_\mu {F^{\mu}}_\nu = 0.
\end{equation}
The $\nu = 3$ equation reads
\begin{equation}
\pd_\mu {F^\mu}_3 = -\frac{\pd^2 A_3}{\pd t^2} = 0,
\end{equation}
whose solutions are given by
\begin{equation}
A_3(t) = A_3(t_0) + C (t-t_0),
\end{equation}
where $C$ is determined by initial conditions. $A_0$ is determined by the fixed-norm constraint $A_0 = \pm \sqrt{A_3^2 - m^2}$. If $C \neq 0$, $A_0$ will eventually evolve to zero. Beyond this point, $A_3$ keeps decreasing, and the fixed-norm condition requires that $A_0$ be imaginary, which is unacceptable since $A_\mu$ is a real-valued vector field. Note that this never happens in the timelike case, as there always exists some real $A_0$ that satisfies the constraint for any value of $A_3$.  The problem is that $A_3$ evolves into the ball $A^2_i < m^2$, which is catastrophic for the spacelike, but not the timelike, case.  An analogous problem arises even when the
Lagrange multiplier constraint is replaced by a smooth potential.

It is possible that this obstruction to a well-defined evolution will be regulated by terms of higher order in the effective field theory.  Using the fixed-norm constraint and solving for $A_0$, the derivative is
\begin{equation}
{\partial_\m A_0} = \frac{A_i}{\sqrt{A_j A_j - m^2}} {\partial_\m A_i }.
\end{equation}
As $A_j A_j$ approaches $m^2$, with finite derivatives of the spatial components, the derivative of the $A_0$ component becomes unbounded. If higher-order terms in the effective action have time derivatives of
the component $A_0$, these terms could become relevant to the vector
field's dynamical evolution, indicating that we have left the realm of
validity of the low-energy effective field theory we are considering.

We are left with the question of how to interpret the timelike Maxwell theory with intial 
data for which $\l \neq 0$.  If  we restrict our attention to initial data for which $\l < 0$ everywhere,
then the evolution of the $A_i$ would be determined and the Hamiltonian would be positive.  We have
\begin{align}
  H &= {1 \over 2} \int d^3 x \,\left( {1 \over 2}F^2_{ij} + (\partial_0 A_i)^2 - (\partial_i A_0)^2 \right)\\
    &= {1 \over 2} \int d^3 x \,\left( {1 \over 2}F^2_{ij} + F_{0i}F_{0i} - 2(\partial_i A_0)F_{i0} \right)\\
    &= {1 \over 2} \int d^3 x \,\left( {1 \over 2}F^2_{ij} + F_{0i}F_{0i} + 2A_0 \partial_i F_{i0} \right)\\     
    &= {1 \over 2} \int d^3 x \,\left( {1 \over 2}F^2_{ij} + F_{0i}F_{0i} - 4 \l A_0^2 \right), \label{ham with lambda}
\end{align}
which is manifestly positive when $ \l < 0 $. However, it is not clear why we should be restricted
to this form of initial data, nor whether even this restriction is enough to ensure stability
beyond perturbation theory.  

The status of this model in both the spacelike and timelike cases remains unclear. However, there are indications of further problems. For the spacelike case, Peloso \emph{et.~al.}~find a linear instability for perturbations with wave numbers on the order of the Hubble parameter in an exponentially expanding cosmology \cite{Himmetoglu:2008zp, Himmetoglu:2008hx}. For the timelike case, Seifert found a gravitational instability in the presence of a spherically symmetric source \cite{Seifert:2007fr}.

\subsection{Scalar action}

The equation of motion for the scalar Lagrangian with a fixed-norm constraint is
\begin{equation}
\partial^\n \partial_\m A^\m = 2 \lambda A^\n.
\end{equation}
Using the fixed-norm constraint ($A_\m A^\m = \mp m^2$), we can solve for the Lagrange multiplier field,
\begin{equation}
\lambda = \mp {1 \over 2 m^2} A_\n \partial^\n \partial_\m A^\m.
\end{equation}
In contrast with the Maxwell theory, in the scalar theory it is the timelike case for which
we can demonstrate obstacles to smooth evolution, while the spacelike case is less clear.
(The Hamiltonian is bounded below, but there are no perturbative instabilities or known
obstacles to smooth evolution.)

When the vector field is timelike, we have four constraint equations in the scalar case,
\begin{equation}
A_0^2 - A_iA_i = m^2  \qquad \text{and} \qquad  \partial_i(\partial_\m A^\m) = 2\lambda A_i.
\end{equation}
Suppose we give homogeneous initial conditions such that $A_0(t_0) > m$.  Align axes such that,
\begin{equation}
A_\m(t_0) = \left(A_0(t_0),0,0,A_3(t_0) \right),
\end{equation}
where $A_3(t_0)^2 = A_0(t_0)^2 - m^2$.  Note that, since $A_3(t_0) \neq 0$, we have that $\lambda = 0$ from the $\n = 3$ equation of motion.  The $\n = 0$ equation of motion therefore gives,
\begin{equation}
{d^2 A_0 \over d t^2} = 0.
\end{equation}
We see that the timelike component of the vector field has the time-evolution, 
\begin{equation}
A_0(t) = A_0(t_0) + C (t-t_0).
\end{equation}  

For generic homogeneous initial conditions,  $C \neq 0$.  In this case, $A_0$ will not have a smooth time evolution since $A_0$ will saturate the fixed-norm constraint, and beyond this point $A_0$ will continue to decrease in magnitude.  To satisfy the fixed-norm constraint, the spatial components of the vector field $A_i$ would need to be imaginary, which is unacceptable since $A_\m$ is a real-valued vector field.  This problem never occurs for the spacelike case since there always exist real values of $A_i$ that satisfy the constraint for any $A_0$.

Again, it is possible that this obstruction to a well-defined evolution will be regulated by terms of higher order in the effective field theory.  The time derivative of $A_3$ is
\begin{equation}
{\partial_\m A_3} = \frac{A_0}{\sqrt{A_0 A_0 - m^2}} {\partial_\m A_0 }.
\end{equation}
As $A_0 A_0$ approaches $m^2$, with finite derivatives of $A_0$, the derivative of the spatial component  $A_3$ becomes unbounded. If higher-order terms in the effective action have time derivatives of
the components $A_i$, these terms could become relevant to the vector
field's dynamical evolution, indicating that we have left the realm of
validity of the low-energy effective field theory we are considering.

Whether or not a theory with a scalar kinetic term and fixed expectation value is viable remains uncertain.

\section{Conclusions}

In this paper, we addressed the issue of stability in theories in which Lorentz invariance is spontaneously broken by a dynamical fixed-norm vector  field with an action
\begin{equation}
S = \int d^4 x \, \left( -{1\over 2}\b_1  F_{\m\n}F^{\m\n} 
-\bnew(\partial_\m A^\m)^2 -\beta_4 {A^\m A^\n \over m^2} (\partial_\m A_\rho)(\partial_\n A^\rho) + \lambda(A^{\mu} A_{\mu} \pm m^2) \right)\,,
\end{equation}
where $\lambda$ is a Lagrange multiplier that strictly enforces the fixed-norm constraint. In the spirit of effective field theory, we limited our attention to only kinetic terms that are quadratic in derivatives, and took care to ensure that our discussion applies to regimes in which an effective field theory expansion is valid.

We examined the boundedness of the Hamiltonian of the theory and showed that, for generic choices of kinetic term, the Hamiltonian is unbounded from below.  Thus for a generic kinetic term, we have shown that a constant fixed-norm background is not the true vacuum of the theory.   The only exception is the timelike sigma-model Lagrangian ($\b_1 = \bnew$, $\beta_4 = 0$ and $A^{\mu} A_{\mu} = -m^2$), in which case the Hamiltonian is positive-definite, ensuring stability.  However, if the vector field instead acquires its vacuum expectation value by minimizing a smooth potential, we demonstrated (as was done previously in \cite{Elliott:2005va}) that the theory is plagued by the existence of a tachyonic ghost, and the Hamiltonian is unbounded from below. 
The timelike fixed-norm sigma-model theory nevertheless serves as a viable starting point for phenomenological investigations of Lorentz invariance; we explore some of this phenomenology in a separate paper \cite{Carroll:2009en}.

We next examined the dispersion relations and energies of first-order perturbations about constant background configurations. We showed that, in addition to the sigma-model case, there are only two other choices of kinetic term for which perturbations have non-negative energies and do not grow exponentially in any frame: the Maxwell ($\bnew = \b_4 = 0$) and scalar ($\b_1 = \b_4 = 0$) Lagrangians. In either case, the theory has fewer than three propagating degrees of freedom
at the linear level, as some of the vector components in the action lack time derivatives and act as additional Lagrange multipliers. 
A subset of the phase space for the Maxwell theory with a timelike \aether\ field is
well-defined and stable, but is identical to ordinary electromagnetism.
For the Maxwell theory with a spacelike \aether\ field, or the scalar theory with a timelike field,
we can find explicit obstructions to smooth time evolution.  It remains unclear whether 
the timelike Maxwell theory or the spacelike scalar theory can exhibit true violation of Lorentz
invariance while remaining well-behaved.

\section*{Acknowledgments} 

We are very grateful to Ted Jacobson, Alan Kostelecky, and Mark Wise for helpful comments.
This research was supported in part by the U.S. Department of Energy and by the 
Gordon and Betty Moore Foundation.


\appendix

\section{Solutions to the linearized equations of motion}\label{Ap:A}

We start by finding the solution to the equations of motion,
linearized about a timelike, fixed-norm background, $A_\m$. Then, showing less details, we find the solutions to the equations of motion linearized about a spacelike background. Finally, we put the solutions in both cases into the compact form of~\eqref{apA:deltaA}-\eqref{scalar mode}.
Our results agree with the solutions for Goldstone modes found in \cite{Gripaios:2004ms}.

The equations of motion for a timelike (+) or spacelike ($-$) vector field are~\eqref{eoms},
\begin{equation}
Q_\m \equiv \left(\eta_{\m \n} \pm {A_\m A_\n \over m^2}\right)\left(\b_1 \pd_\r \pd^\r A^\n + (\bnew-\b_1)\pd^\n \pd_\r A^\r + {\beta_4 } G^\n \right) = 0,
\end{equation} where $G^\n$ is defined in~\eqref{Gnu} and $A^\m Q_\m = 0$ identically.

\paragraph*{Timelike background.} Consider perturbations about an arbitrary, constant (in space and time) timelike background $A_\m = \bar{A}_\m$ that satisfies the constraint: $\bar{A}_\m \bar{A}^\m = - m^2$. Define perturbations by $A_\m = \bA_\m + \d A_\m$. 
Then, to first order in these perturbations, $\bA^\m Q_\m = 0$ identically, and
$\eta^{\m \n} \bA_\m \d A_\n = 0$ by the constraint. 
We can define a basis set of four Lorentz 4-vectors $n^\alpha$, with components
\begin{equation}\label{T vielbeins}
n^0_\m = \bA_\m / m \ , \qquad n^i_\m\ ;\qquad \; i \in \{ 1, 2, 3\}\,,
\end{equation} such that 
\begin{equation}\label{T vielbein ortho}
\eta^{\m \n} n^\a_\m n^\b_\n = \eta^{\a \b}.
\end{equation} 

The independent perturbations are $\d a^\a \equiv \eta^{\m \n} n^\a_\m \d A_\n$ for $\a = 1, 2, 3$. ($\d a^0$ is zero at first order in perturbations due to the constraint.)  It is then clear that there are three independent equations of motion at first order in pertubations (assuming the constraint) for the three independent perturbations,
\begin{equation}
\d Q^i \equiv n^i_\n \left(\b_1 \pd_\r \pd^\r \d A^\n + (\bnew-\b_1)\pd^\n \pd_\r \d A^\r  + \beta_4 n^0_\m n^0_\rho \partial^\m \partial^\rho \delta A^\n \right) = 0,
\end{equation} where $i \in \{ 1, 2, 3\}$. 
We look for plane wave solutions for the $\d A$: 
\begin{equation}\label{T fourier}
\d A_\m = \int d^4 k \, q_\m (k) e^{i k_\n x^\n}.
\end{equation} 
Since $\eta^{\m \n} n^0_\m \d A_\n = 0$, at first order, 
\begin{equation}\label{T q decomp}
q_\m = c_j n^j_\m \qquad \text{where} \qquad j \in \{1,2,3\}.
\end{equation}
The equations of motion become the algebraic equations:
\begin{align}
0	&= \left( \b_1 k_\r k^\r n^i_\n n^{j \n} + (\bnew - \b_1) n^i_\n k^\n n^j_\m k^\m + \beta_4 n^0_\m n^0_\rho k^\m k^\rho n^i_\n n^{j \n} \right) c_j\\
	&= \left( \b_1 k_\r k^\r \d^{i j} + (\bnew - \b_1) n^i_\n k^\n n^j_\m k^\m + \beta_4 n^0_\m n^0_\rho k^\m k^\rho \d^{i j} \right) c_j\\
	&\equiv M^{i j} c_j.
\end{align}

The three independent solutions to these equations are given by
setting an eigenvalue of the matrix $M$ to zero and setting $c_i$ to
the corresponding eigenvector. Setting an eigenvalue of $M$ equal to zero
gives a dispersion relation,
\begin{equation}
\b_1 k_\r k^\r + \beta_4 (n^0_\m k^\m)^2 = 0,
\end{equation} 
with two linearly independent eigenvectors,
\begin{equation}
  \label{eq:speedoflightTmodes}
  (e_2)_i = \e_{2 i j}n^j_\m k^\m \qquad ; \qquad (e_3)_i = \e_{3 i j}n^j_\m k^\m.
\end{equation}
The second eigenvalue of $M$ gives the dispersion relation,  
\begin{align}
\bnew k_\r k^\r + (\bnew - \b_1+\beta_4)(n^0_\m k^\m)^2= 0,\label{weirdTdispersion}
\end{align}
with corresponding eigenvector,
\begin{equation}
  \label{eq:weirdTmodes}
  c_i = n^i_\m k^\m.
\end{equation}

\paragraph*{Spacelike background.} The first order linearized equations of motion about a spacelike background are:
\begin{equation}
\d Q^a \equiv n^a_\n \left(\b_1 \pd_\r \pd^\r \d A^\n + (\bnew-\b_1)\pd^\n \pd_\r \d A^\r + \beta_4 n^3_\m n^3_\rho \partial^\m \partial^\rho \delta A^\n\right) = 0 
\end{equation} where $a \in \{ 0, 1, 2\}$ and where, similarly to the timelike case, we have defined the set of four Lorentz 4-vectors, $n^\a_\m$, to be 
\begin{equation}\label{S vielbeins}
n^3_\m = \bA_\m / m \qquad \text{and} \qquad n^a_\m ; \; a \in \{ 0,1, 2\}
\end{equation} such that 
\begin{equation}\label{S vielbein ortho}
\eta^{\m \n} n^\a_\m n^\b_\n = \eta^{\a \b}.
\end{equation} 
The independent perturbations are $\d a^\a \equiv \eta^{\m \n}
n^\a_\m \d A_\n$ 
for $\a = 0,1,2$. ($\d a^3$ is zero at first order in perturbations
due to the 
constraint.) 

Again we look for plane wave solutions of the form in~\eqref{T fourier}. But now, since $\eta^{\m \n} n^3_\m \d A_\n = 0$, at first order, 
\begin{equation}\label{S q decomp}
q_\m = c_a n^a_\m \qquad \text{where} \qquad a \in \{0,1,2\}.
\end{equation}
The equations of motion become the algebraic equations:
\begin{align}
	&= \left( \b_1 k_\r k^\r n^a_\n n^{b \n} + (\bnew - \b_1) n^a_\n k^\n n^b_\m k^\m + \beta_4 n^3_\m n^3_\rho k^\m k^\rho n^a_\n n^{b \n}\right) c_b\\
	&= \left( \b_1 k_\r k^\r \h^{a b} + (\bnew - \b_1) n^a_\n k^\n n^b_\m k^\m+ \beta_4 n^3_\m n^3_\rho k^\m k^\rho \h^{ab} \right) c_b\\
	&\equiv M^{a b} c_b. \qquad a,b\in\{0,1,2\}
\end{align}
Two independent solutions correspond to the dispersion relation ($a \in \{0,1,2\}$)
\begin{equation}
\b_1 k_\r k^\r + \beta_4 (n^3_\m k^\m)^2 = 0 \,,
\end{equation} 
with corresponding eigenmodes
\begin{equation}
  \label{eq:speedoflightSmodes}
  (e_1)_a = \e_{a 1 b 3}n^b_\m k^\m \qquad ; \qquad (e_2)_a = \e_{a b
    2 3}n^b_\m k^\m.
\end{equation} 
The third solution corresponds to the dispersion relation
\begin{equation}
\label{weirdSdispersion}
\bnew k_\r k^\r - (\bnew - \beta_1-\beta_4)(n^3_\m k^\m)^2= 0\,,
\end{equation}
with corresponding eigenmode
\begin{equation}
  \label{eq:weirdTmodes}
  c_a = \eta_{a b} n^b_\m k^\m.
\end{equation}

\paragraph*{General expression.} We can express the solutions in the timelike and spacelike cases in a compact form by using the
orthonormality of the $n^\a_\m$,~\eqref{T vielbein ortho}, along
with~\eqref{T vielbeins}, \eqref{S vielbeins}, and the fact
that,\footnote{This follows from the invariance of the Levi-Civita tensor,
$$
  \e_{\a \b \g \d}n^\a_\m n^\b_\n n^\g_\r n^\d_\s = \e_{\m \n \r \s}
$$
plus orthonormality,~\eqref{T vielbein ortho}.}
\begin{equation}
  \label{epsilonrelation}
  \e_{\a \b \r \s}n^\a_\m n^\b_\n = \e_{\m \n \a \b}n^\a_\r n^\b_\s.
\end{equation}
Then plugging~\eqref{T q decomp} and \eqref{S q decomp} into~\eqref{T fourier} yields the solutions,
\begin{equation}
\label{apA:deltaA}
\d A_\m = \int d^4 k \, q_\m(k) e^{i k_\n x^\n}
\end{equation} where either,
\begin{equation}
\label{apA:spin-1}
q_\m(k) = i \a^\n k^\r {\bA^\s\over m} \e_{\m \n \r
  \s}~~\text{and}~~\b_1 k_\r k^\r + \beta_4 \left(\bar{A}_\m k^\m \over m\right)^2 = 0~~\text{and}~~\a^\n \bA_\n = 0,
\end{equation} where $\a^\n$ are real-valued constants or,
\begin{equation}
\label{scalar mode}
q_\m =  i \a \left(\h_{\m \n} \pm {\bA_{\m} \bA_{\n}
    \over m^2}\right)k^\n \qquad 
\text{and} \qquad \bnew k_\r k^\r \pm (\bnew-\beta_1 \pm \beta_4) \left(\bar{A}_\m k^\m \over m\right)^2 = 0,
\end{equation}
where $\a$ is a real-valued constant. The reality of the $\a$'s  follows from the condition, $q_\m(k) = q_\m^*(-k)$, that holds if and only if $\d A_\m$ in~\eqref{T fourier} is real. 
In~\eqref{scalar mode},
the ``$+$'' sign corresponds to the timelike background and the
``$-$'' sign to a spacelike background.

\bibliography{lorentz-bib}

\begin{thebibliography}{36}
\expandafter\ifx\csname natexlab\endcsname\relax\def\natexlab#1{#1}\fi
\expandafter\ifx\csname bibnamefont\endcsname\relax
  \def\bibnamefont#1{#1}\fi
\expandafter\ifx\csname bibfnamefont\endcsname\relax
  \def\bibfnamefont#1{#1}\fi
\expandafter\ifx\csname citenamefont\endcsname\relax
  \def\citenamefont#1{#1}\fi
\expandafter\ifx\csname url\endcsname\relax
  \def\url#1{\texttt{#1}}\fi
\expandafter\ifx\csname urlprefix\endcsname\relax\def\urlprefix{URL }\fi
\providecommand{\bibinfo}[2]{#2}
\providecommand{\eprint}[2][]{\url{#2}}

\bibitem[{\citenamefont{Will and Nordtvedt}(1972)}]{Will:1972zz}
\bibinfo{author}{\bibfnamefont{C.~M.} \bibnamefont{Will}} \bibnamefont{and}
  \bibinfo{author}{\bibfnamefont{J.}~\bibnamefont{Nordtvedt},
  \bibfnamefont{Kenneth}}, \bibinfo{journal}{Astrophys. J.}
  \textbf{\bibinfo{volume}{177}}, \bibinfo{pages}{757} (\bibinfo{year}{1972}).

\bibitem[{\citenamefont{Gasperini}(1987)}]{Gasperini:1987nq}
\bibinfo{author}{\bibfnamefont{M.}~\bibnamefont{Gasperini}},
  \bibinfo{journal}{Class. Quant. Grav.} \textbf{\bibinfo{volume}{4}},
  \bibinfo{pages}{485} (\bibinfo{year}{1987}).

\bibitem[{\citenamefont{Kostelecky and Samuel}(1989)}]{Kostelecky:1989jw}
\bibinfo{author}{\bibfnamefont{V.~A.} \bibnamefont{Kostelecky}}
  \bibnamefont{and} \bibinfo{author}{\bibfnamefont{S.}~\bibnamefont{Samuel}},
  \bibinfo{journal}{Phys. Rev.} \textbf{\bibinfo{volume}{D40}},
  \bibinfo{pages}{1886} (\bibinfo{year}{1989}).

\bibitem[{\citenamefont{Colladay and Kostelecky}(1998)}]{Colladay:1998fq}
\bibinfo{author}{\bibfnamefont{D.}~\bibnamefont{Colladay}} \bibnamefont{and}
  \bibinfo{author}{\bibfnamefont{V.~A.} \bibnamefont{Kostelecky}},
  \bibinfo{journal}{Phys. Rev.} \textbf{\bibinfo{volume}{D58}},
  \bibinfo{pages}{116002} (\bibinfo{year}{1998}), \eprint{hep-ph/9809521}.

\bibitem[{\citenamefont{Jacobson and Mattingly}(2001)}]{Jacobson:2000xp}
\bibinfo{author}{\bibfnamefont{T.}~\bibnamefont{Jacobson}} \bibnamefont{and}
  \bibinfo{author}{\bibfnamefont{D.}~\bibnamefont{Mattingly}},
  \bibinfo{journal}{Phys. Rev.} \textbf{\bibinfo{volume}{D64}},
  \bibinfo{pages}{024028} (\bibinfo{year}{2001}), \eprint{gr-qc/0007031}.

\bibitem[{\citenamefont{Eling and Jacobson}(2004)}]{Eling:2003rd}
\bibinfo{author}{\bibfnamefont{C.}~\bibnamefont{Eling}} \bibnamefont{and}
  \bibinfo{author}{\bibfnamefont{T.}~\bibnamefont{Jacobson}},
  \bibinfo{journal}{Phys. Rev.} \textbf{\bibinfo{volume}{D69}},
  \bibinfo{pages}{064005} (\bibinfo{year}{2004}), \eprint{gr-qc/0310044}.

\bibitem[{\citenamefont{Carroll and Lim}(2004)}]{Carroll:2004ai}
\bibinfo{author}{\bibfnamefont{S.~M.} \bibnamefont{Carroll}} \bibnamefont{and}
  \bibinfo{author}{\bibfnamefont{E.~A.} \bibnamefont{Lim}},
  \bibinfo{journal}{Phys. Rev.} \textbf{\bibinfo{volume}{D70}},
  \bibinfo{pages}{123525} (\bibinfo{year}{2004}), \eprint{hep-th/0407149}.

\bibitem[{\citenamefont{Jacobson and Mattingly}(2004)}]{Jacobson:2004ts}
\bibinfo{author}{\bibfnamefont{T.}~\bibnamefont{Jacobson}} \bibnamefont{and}
  \bibinfo{author}{\bibfnamefont{D.}~\bibnamefont{Mattingly}},
  \bibinfo{journal}{Phys. Rev.} \textbf{\bibinfo{volume}{D70}},
  \bibinfo{pages}{024003} (\bibinfo{year}{2004}), \eprint{gr-qc/0402005}.

\bibitem[{\citenamefont{Lim}(2005)}]{Lim:2004js}
\bibinfo{author}{\bibfnamefont{E.~A.} \bibnamefont{Lim}},
  \bibinfo{journal}{Phys. Rev.} \textbf{\bibinfo{volume}{D71}},
  \bibinfo{pages}{063504} (\bibinfo{year}{2005}), \eprint{astro-ph/0407437}.

\bibitem[{\citenamefont{Eling et~al.}(2004)\citenamefont{Eling, Jacobson, and
  Mattingly}}]{Eling:2004dk}
\bibinfo{author}{\bibfnamefont{C.}~\bibnamefont{Eling}},
  \bibinfo{author}{\bibfnamefont{T.}~\bibnamefont{Jacobson}}, \bibnamefont{and}
  \bibinfo{author}{\bibfnamefont{D.}~\bibnamefont{Mattingly}}
  (\bibinfo{year}{2004}), \eprint{gr-qc/0410001}.

\bibitem[{\citenamefont{Dulaney et~al.}(2008)\citenamefont{Dulaney, Gresham,
  and Wise}}]{Dulaney:2008ph}
\bibinfo{author}{\bibfnamefont{T.~R.} \bibnamefont{Dulaney}},
  \bibinfo{author}{\bibfnamefont{M.~I.} \bibnamefont{Gresham}},
  \bibnamefont{and} \bibinfo{author}{\bibfnamefont{M.~B.} \bibnamefont{Wise}},
  \bibinfo{journal}{Phys. Rev.} \textbf{\bibinfo{volume}{D77}},
  \bibinfo{pages}{083510} (\bibinfo{year}{2008}), \eprint{0801.2950}.

\bibitem[{\citenamefont{Jimenez and Maroto}(2008)}]{Jimenez:2008sq}
\bibinfo{author}{\bibfnamefont{J.~B.} \bibnamefont{Jimenez}} \bibnamefont{and}
  \bibinfo{author}{\bibfnamefont{A.~L.} \bibnamefont{Maroto}}
  (\bibinfo{year}{2008}), \eprint{0811.0784}.

\bibitem[{\citenamefont{Kostelecky and Lehnert}(2001)}]{Kostelecky:2000mm}
\bibinfo{author}{\bibfnamefont{V.~A.} \bibnamefont{Kostelecky}}
  \bibnamefont{and} \bibinfo{author}{\bibfnamefont{R.}~\bibnamefont{Lehnert}},
  \bibinfo{journal}{Phys. Rev.} \textbf{\bibinfo{volume}{D63}},
  \bibinfo{pages}{065008} (\bibinfo{year}{2001}), \eprint{hep-th/0012060}.

\bibitem[{\citenamefont{Elliott et~al.}(2005)\citenamefont{Elliott, Moore, and
  Stoica}}]{Elliott:2005va}
\bibinfo{author}{\bibfnamefont{J.~W.} \bibnamefont{Elliott}},
  \bibinfo{author}{\bibfnamefont{G.~D.} \bibnamefont{Moore}}, \bibnamefont{and}
  \bibinfo{author}{\bibfnamefont{H.}~\bibnamefont{Stoica}},
  \bibinfo{journal}{JHEP} \textbf{\bibinfo{volume}{08}}, \bibinfo{pages}{066}
  (\bibinfo{year}{2005}), \eprint{hep-ph/0505211}.

\bibitem[{\citenamefont{Mattingly}(2005)}]{Mattingly:2005re}
\bibinfo{author}{\bibfnamefont{D.}~\bibnamefont{Mattingly}},
  \bibinfo{journal}{Living Rev. Rel.} \textbf{\bibinfo{volume}{8}},
  \bibinfo{pages}{5} (\bibinfo{year}{2005}), \eprint{gr-qc/0502097}.

\bibitem[{\citenamefont{Will}(2005)}]{Will:2005va}
\bibinfo{author}{\bibfnamefont{C.~M.} \bibnamefont{Will}},
  \bibinfo{journal}{Living Rev. Rel.} \textbf{\bibinfo{volume}{9}},
  \bibinfo{pages}{3} (\bibinfo{year}{2005}), \eprint{gr-qc/0510072}.

\bibitem[{\citenamefont{Jacobson}(2008)}]{Jacobson:2008aj}
\bibinfo{author}{\bibfnamefont{T.}~\bibnamefont{Jacobson}}
  (\bibinfo{year}{2008}), \eprint{0801.1547}.

\bibitem[{\citenamefont{Arkani-Hamed et~al.}(2004)\citenamefont{Arkani-Hamed,
  Cheng, Luty, and Mukohyama}}]{ArkaniHamed:2003uy}
\bibinfo{author}{\bibfnamefont{N.}~\bibnamefont{Arkani-Hamed}},
  \bibinfo{author}{\bibfnamefont{H.-C.} \bibnamefont{Cheng}},
  \bibinfo{author}{\bibfnamefont{M.~A.} \bibnamefont{Luty}}, \bibnamefont{and}
  \bibinfo{author}{\bibfnamefont{S.}~\bibnamefont{Mukohyama}},
  \bibinfo{journal}{JHEP} \textbf{\bibinfo{volume}{05}}, \bibinfo{pages}{074}
  (\bibinfo{year}{2004}), \eprint{hep-th/0312099}.

\bibitem[{\citenamefont{Arkani-Hamed et~al.}(2007)\citenamefont{Arkani-Hamed,
  Cheng, Luty, Mukohyama, and Wiseman}}]{ArkaniHamed:2005gu}
\bibinfo{author}{\bibfnamefont{N.}~\bibnamefont{Arkani-Hamed}},
  \bibinfo{author}{\bibfnamefont{H.-C.} \bibnamefont{Cheng}},
  \bibinfo{author}{\bibfnamefont{M.~A.} \bibnamefont{Luty}},
  \bibinfo{author}{\bibfnamefont{S.}~\bibnamefont{Mukohyama}},
  \bibnamefont{and} \bibinfo{author}{\bibfnamefont{T.}~\bibnamefont{Wiseman}},
  \bibinfo{journal}{JHEP} \textbf{\bibinfo{volume}{01}}, \bibinfo{pages}{036}
  (\bibinfo{year}{2007}), \eprint{hep-ph/0507120}.

\bibitem[{\citenamefont{Cheng et~al.}(2006)\citenamefont{Cheng, Luty,
  Mukohyama, and Thaler}}]{Cheng:2006us}
\bibinfo{author}{\bibfnamefont{H.-C.} \bibnamefont{Cheng}},
  \bibinfo{author}{\bibfnamefont{M.~A.} \bibnamefont{Luty}},
  \bibinfo{author}{\bibfnamefont{S.}~\bibnamefont{Mukohyama}},
  \bibnamefont{and} \bibinfo{author}{\bibfnamefont{J.}~\bibnamefont{Thaler}},
  \bibinfo{journal}{JHEP} \textbf{\bibinfo{volume}{05}}, \bibinfo{pages}{076}
  (\bibinfo{year}{2006}), \eprint{hep-th/0603010}.

\bibitem[{\citenamefont{Carroll et~al.}(2003)\citenamefont{Carroll, Hoffman,
  and Trodden}}]{Carroll:2003st}
\bibinfo{author}{\bibfnamefont{S.~M.} \bibnamefont{Carroll}},
  \bibinfo{author}{\bibfnamefont{M.}~\bibnamefont{Hoffman}}, \bibnamefont{and}
  \bibinfo{author}{\bibfnamefont{M.}~\bibnamefont{Trodden}},
  \bibinfo{journal}{Phys. Rev.} \textbf{\bibinfo{volume}{D68}},
  \bibinfo{pages}{023509} (\bibinfo{year}{2003}), \eprint{astro-ph/0301273}.

\bibitem[{\citenamefont{Carroll et~al.}(2009)\citenamefont{Carroll, Dulaney,
  Gresham, and Tam}}]{Carroll:2009en}
\bibinfo{author}{\bibfnamefont{S.~M.} \bibnamefont{Carroll}},
  \bibinfo{author}{\bibfnamefont{T.~R.} \bibnamefont{Dulaney}},
  \bibinfo{author}{\bibfnamefont{M.~I.} \bibnamefont{Gresham}},
  \bibnamefont{and} \bibinfo{author}{\bibfnamefont{H.}~\bibnamefont{Tam}},
  \bibinfo{journal}{Phys. Rev.} \textbf{\bibinfo{volume}{D79}},
  \bibinfo{pages}{065012} (\bibinfo{year}{2009}), \eprint{0812.1050}.

\bibitem[{\citenamefont{Dubovsky and Sibiryakov}(2006)}]{Dubovsky:2006vk}
\bibinfo{author}{\bibfnamefont{S.~L.} \bibnamefont{Dubovsky}} \bibnamefont{and}
  \bibinfo{author}{\bibfnamefont{S.~M.} \bibnamefont{Sibiryakov}},
  \bibinfo{journal}{Phys. Lett.} \textbf{\bibinfo{volume}{B638}},
  \bibinfo{pages}{509} (\bibinfo{year}{2006}), \eprint{hep-th/0603158}.

\bibitem[{\citenamefont{Eling et~al.}(2007)\citenamefont{Eling, Foster,
  Jacobson, and Wall}}]{Eling:2007qd}
\bibinfo{author}{\bibfnamefont{C.}~\bibnamefont{Eling}},
  \bibinfo{author}{\bibfnamefont{B.~Z.} \bibnamefont{Foster}},
  \bibinfo{author}{\bibfnamefont{T.}~\bibnamefont{Jacobson}}, \bibnamefont{and}
  \bibinfo{author}{\bibfnamefont{A.~C.} \bibnamefont{Wall}},
  \bibinfo{journal}{Phys. Rev.} \textbf{\bibinfo{volume}{D75}},
  \bibinfo{pages}{101502} (\bibinfo{year}{2007}), \eprint{hep-th/0702124}.

\bibitem[{\citenamefont{Kostelecky}(2001)}]{Kostelecky:2001xz}
\bibinfo{author}{\bibfnamefont{V.~A.} \bibnamefont{Kostelecky}}
  (\bibinfo{year}{2001}), \eprint{hep-ph/0104227}.

\bibitem[{\citenamefont{Adams et~al.}(2006)\citenamefont{Adams, Arkani-Hamed,
  Dubovsky, Nicolis, and Rattazzi}}]{Adams:2006sv}
\bibinfo{author}{\bibfnamefont{A.}~\bibnamefont{Adams}},
  \bibinfo{author}{\bibfnamefont{N.}~\bibnamefont{Arkani-Hamed}},
  \bibinfo{author}{\bibfnamefont{S.}~\bibnamefont{Dubovsky}},
  \bibinfo{author}{\bibfnamefont{A.}~\bibnamefont{Nicolis}}, \bibnamefont{and}
  \bibinfo{author}{\bibfnamefont{R.}~\bibnamefont{Rattazzi}},
  \bibinfo{journal}{JHEP} \textbf{\bibinfo{volume}{10}}, \bibinfo{pages}{014}
  (\bibinfo{year}{2006}), \eprint{hep-th/0602178}.

\bibitem[{\citenamefont{Bluhm et~al.}(2008{\natexlab{a}})\citenamefont{Bluhm,
  Gagne, Potting, and Vrublevskis}}]{Bluhm:2008yt}
\bibinfo{author}{\bibfnamefont{R.}~\bibnamefont{Bluhm}},
  \bibinfo{author}{\bibfnamefont{N.~L.} \bibnamefont{Gagne}},
  \bibinfo{author}{\bibfnamefont{R.}~\bibnamefont{Potting}}, \bibnamefont{and}
  \bibinfo{author}{\bibfnamefont{A.}~\bibnamefont{Vrublevskis}},
  \bibinfo{journal}{Phys. Rev.} \textbf{\bibinfo{volume}{D77}},
  \bibinfo{pages}{125007} (\bibinfo{year}{2008}{\natexlab{a}}),
  \eprint{0802.4071}.

\bibitem[{\citenamefont{Chkareuli et~al.}(2006)\citenamefont{Chkareuli,
  Froggatt, and Nielsen}}]{Chkareuli:2006yf}
\bibinfo{author}{\bibfnamefont{J.~L.} \bibnamefont{Chkareuli}},
  \bibinfo{author}{\bibfnamefont{C.~D.} \bibnamefont{Froggatt}},
  \bibnamefont{and} \bibinfo{author}{\bibfnamefont{H.~B.}
  \bibnamefont{Nielsen}} (\bibinfo{year}{2006}), \eprint{hep-th/0610186}.

\bibitem[{\citenamefont{Gripaios}(2004)}]{Gripaios:2004ms}
\bibinfo{author}{\bibfnamefont{B.~M.} \bibnamefont{Gripaios}},
  \bibinfo{journal}{JHEP} \textbf{\bibinfo{volume}{10}}, \bibinfo{pages}{069}
  (\bibinfo{year}{2004}), \eprint{hep-th/0408127}.

\bibitem[{\citenamefont{Clayton}(2001)}]{Clayton:2001vy}
\bibinfo{author}{\bibfnamefont{M.~A.} \bibnamefont{Clayton}}
  (\bibinfo{year}{2001}), \eprint{gr-qc/0104103}.

\bibitem[{\citenamefont{Henneaux and Teitelboim}(1992)}]{Henneaux:1992ir}
\bibinfo{author}{\bibfnamefont{M.}~\bibnamefont{Henneaux}} \bibnamefont{and}
  \bibinfo{author}{\bibfnamefont{C.}~\bibnamefont{Teitelboim}},
  \emph{\bibinfo{title}{Quantization of gauge systems}}
  (\bibinfo{publisher}{Princeton University}, \bibinfo{address}{Princeton,
  N.J.}, \bibinfo{year}{1992}), ISBN \bibinfo{isbn}{069108775X (acid-free
  paper)},
  \urlprefix\url{http://www.loc.gov/catdir/description/prin031/92011585.html}.

\bibitem[{\citenamefont{Nambu}(1968)}]{Nambu:1968qk}
\bibinfo{author}{\bibfnamefont{Y.}~\bibnamefont{Nambu}},
  \bibinfo{journal}{Progr. Theoret. Phys. Suppl. Extra No.} pp.
  \bibinfo{pages}{190--195} (\bibinfo{year}{1968}).

\bibitem[{\citenamefont{Bluhm et~al.}(2008{\natexlab{b}})\citenamefont{Bluhm,
  Fung, and Kostelecky}}]{Bluhm:2007bd}
\bibinfo{author}{\bibfnamefont{R.}~\bibnamefont{Bluhm}},
  \bibinfo{author}{\bibfnamefont{S.-H.} \bibnamefont{Fung}}, \bibnamefont{and}
  \bibinfo{author}{\bibfnamefont{V.~A.} \bibnamefont{Kostelecky}},
  \bibinfo{journal}{Phys. Rev.} \textbf{\bibinfo{volume}{D77}},
  \bibinfo{pages}{065020} (\bibinfo{year}{2008}{\natexlab{b}}),
  \eprint{0712.4119}.

\bibitem[{\citenamefont{Himmetoglu
  et~al.}(2008{\natexlab{a}})\citenamefont{Himmetoglu, Contaldi, and
  Peloso}}]{Himmetoglu:2008zp}
\bibinfo{author}{\bibfnamefont{B.}~\bibnamefont{Himmetoglu}},
  \bibinfo{author}{\bibfnamefont{C.~R.} \bibnamefont{Contaldi}},
  \bibnamefont{and} \bibinfo{author}{\bibfnamefont{M.}~\bibnamefont{Peloso}}
  (\bibinfo{year}{2008}{\natexlab{a}}), \eprint{0809.2779}.

\bibitem[{\citenamefont{Himmetoglu
  et~al.}(2008{\natexlab{b}})\citenamefont{Himmetoglu, Contaldi, and
  Peloso}}]{Himmetoglu:2008hx}
\bibinfo{author}{\bibfnamefont{B.}~\bibnamefont{Himmetoglu}},
  \bibinfo{author}{\bibfnamefont{C.~R.} \bibnamefont{Contaldi}},
  \bibnamefont{and} \bibinfo{author}{\bibfnamefont{M.}~\bibnamefont{Peloso}}
  (\bibinfo{year}{2008}{\natexlab{b}}), \eprint{0812.1231}.

\bibitem[{\citenamefont{Seifert}(2007)}]{Seifert:2007fr}
\bibinfo{author}{\bibfnamefont{M.~D.} \bibnamefont{Seifert}},
  \bibinfo{journal}{Phys. Rev.} \textbf{\bibinfo{volume}{D76}},
  \bibinfo{pages}{064002} (\bibinfo{year}{2007}), \eprint{gr-qc/0703060}.

\end{thebibliography}

\end{document}